\documentclass[final,5p,times]{elsarticle}
\AtBeginDocument{\RenewCommandCopy\qty\SI}
\usepackage{amssymb,bbold}
\usepackage{amsmath}
\usepackage[utf8x]{inputenc}
\usepackage{natbib}
\usepackage{graphicx}
\usepackage{hyperref}
\usepackage{epstopdf}
\usepackage{pifont}
\usepackage{float}
\usepackage{ulem}
\usepackage{cuted}
\usepackage{physics}
\usepackage{soul}
\usepackage{cancel}
\usepackage{orcidlink}
\usepackage{cuted}
\usepackage{multirow}

\usepackage{booktabs,multirow,siunitx}
\graphicspath{{pdf/}}
\hypersetup{colorlinks=true,linkcolor=blue,citecolor=blue,filecolor=blue,urlcolor=blue,breaklinks=true}
\biboptions{sort&compress}
\journal{Annals of Physics}
\RequirePackage{color}

\begin{document}

\begin{frontmatter}

\title{Nonlinear optical behavior of confined electrons under torsion and magnetic fields}

% --- Rótulos distintos para cada afiliação ---
\author[ppgf]{Carlos Magno O. Pereira\,\orcidlink{0000-0002-5170-7538}}
\ead{carlos.mop@discente.ufma.br}

\author[ppgf,df]{Edilberto O. Silva\,\orcidlink{0000-0002-0297-5747}}
\ead{edilberto.silva@ufma.br}

% --- Definição de cada endereço único uma só vez ---
\address[ppgf]{Programa de P\'{o}s-Gradua\c{c}\~{a}o em F\'{i}sica, Universidade Federal do Maranh\~{a}o, Campus Universit\'{a}rio do Bacanga, 65080-805, S\~{a}o Lu\'{i}s, Maranh\~{a}o, Brazil}

\address[df]{Coordena\c c\~ao do Curso de F\'{\i}sica -- Bacharelado, Universidade Federal do Maranh\~{a}o, Campus Universit\'{a}rio do Bacanga, 65085-580 S\~{a}o Lu\'{\i}s, Maranh\~{a}o, Brazil}

\begin{abstract}
In this work, we investigate the influence of torsion, Aharonov-Bohm flux, and external magnetic fields on the linear and nonlinear optical properties of a confined quantum system. The confinement potential is not assumed a priori, but emerges as a radial effective potential, analogous to a quantum dot, geometrically induced by the torsion of the material. Starting from an effective radial equation derived in a nontrivial geometric background, we analytically solve for the energy spectrum and wave functions. These solutions are then employed to evaluate the optical absorption coefficients and refractive index changes, including both linear and third-order nonlinear contributions. The formalism incorporates the electric dipole approximation and accounts for intensity-dependent effects such as saturation and spectral shifts. Our results reveal that torsion and topological parameters significantly modify the optical response, leading to tunable resonances and nontrivial dispersive behavior. This work highlights the potential of geometric and topological engineering in low-dimensional systems to control and enhance nonlinear optical phenomena.
\end{abstract}


\end{frontmatter}

\section{Introduction}
\label{sec:intro}

Low-dimensional materials (LDMs) and nanostructures provide a fertile platform to explore quantum phenomena and to engineer tailored electronic and optical responses at the nanoscale \cite{RevModPhys.88.025005,PhysRevB.40.1620,PhysRevB.80.115316,PhysRevB.82.115210,PhysRevB.110.134513,LSA.2024.13.30,MRL.2023.11.21}. Quantum confinement in quantum wells, nanowires, quantum dots (QDs), and quantum rings (QRs) reshapes the density of states and selection rules, enabling photodetectors, modulators, and elements for integrated photonics \cite{Levine1993QWIP,Miller1984QCSE_PRL,Miller1986QCSE_Devices,Konstantatos2010CQD_PD_Review,Rogalski2002QDIP_Review,Yan2009NanowirePhotonics,Arakawa1982QDlaser_APL,Wang2019QD_SiPhot,Lorke2000QR_PRL,Warburton2000QD_Nature}
. In parallel, van der Waals heterostructures and strain engineering offer additional parameters, layer stacking, twist, and mechanical deformation, to tune band structures, excitonic resonances, and nonlinear susceptibilities \cite{YU2025130856,10.1063/5.0037852,PhysRevB.109.094105,PhysRevB.109.115434,Li2024}.

Beyond band-structure and potential-shape engineering, geometry and topology have emerged as powerful organizing principles for quantum states. The geometric theory of defects models crystalline dislocations by non-Euclidean metrics, in which torsion (but not necessarily curvature) modifies the Laplace--Beltrami operator that governs kinetic energy \cite{Katanaev1992AnnPhys,Kleinert1989GaugeFields,Nelson1987DefectsGeometry,Jensen1991,Puntigam1997CQG,Katanaev2005PhysUsp,Furtado1999PLA,Moraes2000IJMPA}. A continuous and homogeneous distribution of screw dislocations can thus be encoded by a cylindrically symmetric metric with uniform torsion, introducing an effective confinement even in the absence of a traditional potential. In mesoscopic rings and dots, this geometric confinement interplays with magnetic and topological phases to reshape spectra and dipole matrix elements, directly impacting observable optical coefficients \cite{Buttiker1983PLA,Gefen1984PRL,Webb1985PRL,Mailly1993PRL,Lorke2000PRL,Reimann2002RMP,Viefers2004PhysE,Sheng2002PRB,Fomin2003PRB,Fomin2014Book}.

A prototypical topological ingredient is the Aharonov--Bohm (AB) effect, whereby the electromagnetic vector potential imprints a quantum phase in multiply connected geometries and modulates energies, persistent currents, and oscillator strengths even when the local magnetic field vanishes \cite{Aharonov1959PR,Aharonov1961PR,Chambers1960PRL,Tonomura1986Nature,Webb1985PRL,Washburn1992RPP,Levy1990PRL,Chandrasekhar1991PRL,Gefen1984PRL,Buttiker1983PLA,Cheung1988PRB,Lorke2000PRL}. When a uniform perpendicular magnetic field is present, electronic motion condenses into Landau levels; in rotationally symmetric confinements (e.g., parabolic dots/rings) this evolves into the Fock--Darwin ladder, whose level spacings and degeneracies are widely used to interpret magneto-transport and magneto-optical experiments \cite{Landau1930,Fock1928,Darwin1930,Maksym1990PRL,McEuen1991PRL,Tarucha1996PRL,Reimann2002RMP,Beenakker1991SSP}. In the setting studied here, the AB phase, magnetic quantization, and torsion act cooperatively: the AB flux controls the angular-momentum content and phase coherence; the field sets the magnetic length and cyclotron scale; and torsion provides a geometric confining energy that compresses wavefunctions radially, simultaneously increasing level spacings and suppressing radial overlaps.

Optical probes are exquisitely sensitive to these ingredients. Linear absorption peaks track inter-level spacings and broadening, while third-order (Kerr-like) nonlinearities encode intensity-dependent saturation and dispersive reshaping through the detuning structure of $\chi^{(3)}$ \cite{SheikBahae1990OL,SheikBahae1991JOSAB,Wherrett1984JOSAB,ChemlaShah2001Nature,Dinu2003APL,Autere2018AdvMater,Eggleton2011NatPhoton,Kippenberg2018Science}. In mesoscopic systems, geometric confinement and topological phases can produce pronounced blueshifts of resonances (via level-spacing enhancement) and strong amplitude suppression (via reduced wave-function overlap), and they may even drive optical switching when the negative third-order contribution exceeds the linear absorption at resonance \cite{Lorke2000PRL,Hu2001PRB,Govorov2002PRB,Miller2010NP}. Complementarily, the photoionization cross-section (PCS) of bound states provides a window onto bound-to-continuum dynamics under fields and topology, with clear signatures in the resonance position and height as AB flux, magnetic field, and geometry are varied \cite{Bastard1990PRB_PCS_QW,Bryant1988PRB_QD_Impurity,Sari1996PRB_QD_PCS,Ishikawa1995PRB_QW_PCS,Szeftel1992PRB_QD_Acceptor_PCS,Chakraborty1994PRB_Rings_AB_PCS,Govorov2004PRB_Rings_Exciton_AB,Maksym1990PRB_FD_Optics,Lugo2006PRB_QD_PCS_Field,Xie2013SSC_QR_PCS,Zhang2010JAP_QD_PCS_SizeField,Sadeghi2011PLA_QR_PCS_AB,Jin2018SciRep_QD_PCS_Exp,Kuyucu2014SST_QD_PCS_Review,Bai2021Nanophotonics_PCS_Topology}.

In this work, we investigate a nonrelativistic electron with effective mass $m^{*}$ confined to a medium carrying a continuous torsion density and subject to both a uniform magnetic field and an AB flux. Starting from the minimally coupled Schrödinger equation in a torsion-bearing metric, we derive the exact radial equation and cast it into a Schrödinger form with an explicit effective potential. Analytical solutions are obtained in terms of confluent hypergeometric (Kummer) functions, yielding normalized radial wavefunctions and closed-form energy spectra that make transparent the roles of torsion, AB phase, and magnetic quantization. These eigenstates are then used to evaluate (i) linear and third-order nonlinear optical absorption and refractive-index changes within the electric-dipole approximation and (ii) the PCS for a representative dipole-allowed transition. We find that torsion systematically blueshifts resonances and suppresses their amplitudes by reducing radial overlaps, and that at high intensities, the system can exhibit optical switching controlled by the combined action of torsion and AB flux. Here, we show that uniform torsion acts as a geometric, non-magnetic control parameter that lifts degeneracies and increases level spacings in ring-like confinements; we quantify how torsion and AB flux reshape linear and third-order optical responses through modified detunings and dipole matrix elements; and we demonstrate that both photoionization resonances and their amplitudes can be tuned by torsion and flux, offering clear experimental signatures of geometric/topological engineering. 

The work is organized as follows. Section~\ref{sec:theoretical_model} introduces the torsion-bearing metric, the minimal-coupling Hamiltonian with AB flux and a uniform magnetic field, and derives the exact radial equation along with its Schrödinger-form effective potential. Section~\ref{sec:optical_properties_framework} develops the optical framework for linear and third-order nonlinear absorption, refractive-index changes, and the PCS formalism with finite broadening. Section~\ref{sec:results} presents numerical results for energies, normalized wavefunctions, absorption, refractive index changes, and PCS, highlighting torsion- and flux-induced blueshifts and amplitude suppression. Section~\ref{sec:osc_strength} details the calculation of the oscillator strength, analyzing its dependence on geometric and topological parameters. Section~\ref{conclusion} closes with the main conclusions, limitations, and possible extensions to include spin-orbit coupling, strain fields, and higher-order or time-resolved nonlinear spectroscopies.

\section{Theoretical Model}
\label{sec:theoretical_model}

We investigate the quantum dynamics of an electron with effective mass $m^{*}$ confined to a medium with a continuous and homogeneous distribution of parallel screw dislocations. In the geometric theory of defects, such a medium is described by a non-Euclidean metric. Following the approach for a cylindrically symmetric distribution, the line element in cylindrical coordinates $(\rho,\varphi,z)$ is given by \cite{BJP.2011.41.167,EPJB.2013.86.315,SilvaNetto2008JPCM,PLA.2012.376.2838} 
\begin{equation}
 ds^2 = \bigl(dz + \tau\rho^2\, d\varphi\bigr)^2 + d\rho^2 + \rho^2\, d\varphi^2,
 \label{eq:metric}
\end{equation}
where $\tau$ is the torsion density parameter, proportional to the surface density of the Burgers vectors of the dislocations. This metric describes a space with uniform torsion and zero curvature.

In addition to the geometric effects, the electron interacts with an external magnetic field. Its corresponding total vector potential, $\boldsymbol{A}$, consists of two components. First, a uniform magnetic field $\boldsymbol{B}=B\,\hat{z}$ is applied along the axis of the defect; the corresponding vector potential in the symmetric gauge is $\boldsymbol{A}_B=(B\rho/2)\,\hat{\varphi}$. Second, we introduce an AB magnetic flux $\Phi$, confined to an infinitesimally thin solenoid along the $z$-axis, with vector potential $\boldsymbol{A}_{\text{AB}}=(\Phi/2\pi\rho)\,\hat{\varphi}$. The total vector potential experienced by the electron is $\boldsymbol{A}=\boldsymbol{A}_B+\boldsymbol{A}_{\text{AB}}$ \cite{AP.2023.459.169547,PRD.2020.102.105020,PLA.2015.379.907,PLA.2016.380.3847,FBS.2022.63.58,PE.2021.132.114760,Pereira2024QuantumReports,AdP.2019.531.1900254,CTP.2018.70.817,EPJC.2015.75.321,Universe.2020.6.203}.

\begin{figure*}[tbhp]
    \centering
    \includegraphics[width=0.9\linewidth]{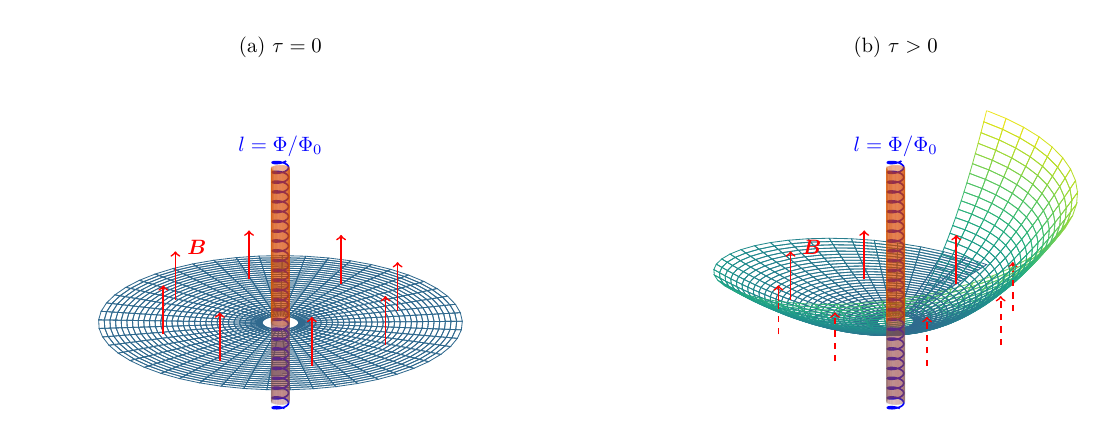}
    \caption{Schematic illustration of the effect of torsion density ($\tau$) on the geometry of the system. (a) The case without torsion ($\tau=0$) corresponds to a two-dimensional flat disk representing Euclidean space. (b) A finite torsion ($\tau>0$) introduces a helical distortion in the surface, resulting in a non-Euclidean geometry. In both configurations, the system is subjected to a uniform magnetic field $\boldsymbol{B}$, applied perpendicular to the plane (indicated by the red arrows), and an Aharonov--Bohm flux $l$, confined along the central axis.}
    \label{fig:torsion}
\end{figure*}

The torsion-induced modification of the geometry gives rise to an effective harmonic confinement potential resulting exclusively from the new spatial configuration. Fig.~\ref{fig:torsion} illustrates the effect of torsion density ($\tau$) on the geometry of a two-dimensional system. Figure~\ref{fig:torsion}(a) shows the reference case, with $\tau=0$, in which the system corresponds to a flat disk. In contrast, Fig.~\ref{fig:torsion}(b) shows that a finite torsion density ($\tau>0$) produces a helical distortion on the surface, resulting in a non-Euclidean geometry and, consequently, an effective confinement potential. In both cases, the system is subjected to a uniform magnetic field $\boldsymbol{B}$, oriented perpendicular to the plane (red arrows), as well as an AB magnetic flux ($l=\Phi/\Phi_0$), confined along the central axis.

In this background, the quantum dynamics of a charged particle are described by the Schr\"odinger equation with minimal coupling. The Hamiltonian is constructed from the Laplace--Beltrami operator for the metric in Eq.~\eqref{eq:metric}, including the interaction with the vector potential $\boldsymbol{A}$
\begin{equation}
 H=\frac{1}{2 m^*}\left(-i\hbar\nabla - e\boldsymbol{A}\right)^2.
\end{equation}
In this non-Euclidean space, the explicit form of the Hamiltonian operator is \cite{JPCM.2008.20.125209}
\begin{equation}
 H=-\frac{\hbar^2}{2 m^*}\left[\,\frac{1}{\rho}\frac{\partial}{\partial\rho}\left(\rho\frac{\partial}{\partial\rho}\right)
 + \left(\frac{1}{\rho}\frac{\partial}{\partial\varphi} - \rho\tau\frac{\partial}{\partial z} - \frac{ieA_\varphi}{\hbar}\right)^{2}
 + \frac{\partial^2}{\partial z^2}\,\right],
 \label{eq:hamiltonian_full}
\end{equation}
where $A_\varphi=B\rho/2+\Phi/(2\pi\rho)$ is the only non-zero component of the total vector potential.

Due to the cylindrical symmetry of the problem, the wave function can be separated as
\begin{equation}
 \Psi(\rho,\varphi,z)=R(\rho)\,e^{im\varphi}\,e^{ik_z z},
 \label{eq:ansatz}
\end{equation}
where $m$ is the integer magnetic quantum number and $k_z$ is the continuous wave number for the free motion along the $z$-axis. Substituting the ansatz from Eq.~\eqref{eq:ansatz} into the time-independent Schr\"odinger equation, $H\Psi=E\Psi$, we obtain the following radial equation for $R(\rho)$:
\begin{align}
 \frac{1}{\rho}\frac{d}{d\rho}\left(\rho\frac{dR}{d\rho}\right)
 &- \frac{(m-l)^2}{\rho^2}\,R - \Lambda^2\rho^2 R
 \notag \\&+ \left[\frac{2m^{*} E}{\hbar^2} - k_z^2 + 2(m-l)\Lambda \right]R = 0,
 \label{eq:radial_equation_initial}
\end{align}
where $l=\Phi/\Phi_0=e\Phi/(2\pi\hbar)$ is the AB flux normalized by the magnetic flux quantum $\Phi_0$, and $\Lambda = k_z \tau + eB/(2\hbar)$ is the characteristic inverse-length-squared factor.

In our study, it is important to derive the expression for the effective potential of the problem.
Starting from Eq.~(\ref{eq:radial_equation_initial}), we introduce the Liouville substitution
\begin{equation}
R(\rho)=\frac{u(\rho)}{\sqrt{\rho}}.
\label{eq:liouville_subst}
\end{equation}
Using the identity
\begin{equation}
\frac{1}{\rho}\frac{d}{d\rho}\left(\rho\frac{d}{d\rho}\frac{u}{\sqrt{\rho}}\right)
=\frac{1}{\sqrt{\rho}}\left(u''+\frac{u}{4\rho^2}\right),
\label{eq:identity}
\end{equation}
the radial equation reduces to the Schr\"odinger form 
\begin{equation}
-\frac{\hbar^2}{2m^*}\,u''(\rho)\;+\;V_{\mathrm{eff}}(\rho)\,u(\rho)\;=\;E\,u(\rho),
\label{eq:schr_pureE}
\end{equation}
where the effective potential is
\begin{equation}
V_{\mathrm{eff}}(\rho)
=\frac{\hbar^2}{2m^*}\left[\frac{(m-l)^2-\tfrac14}{\rho^2}
+\Lambda^2\rho^2\right]
-\frac{\hbar^2}{m^*}(m-l)\Lambda
+\frac{\hbar^2 k_z^2}{2m^*}.
\label{eq:Veff_pureE}
\end{equation}

The physical origin of the harmonic confinement term  $(\propto \Lambda^2 \rho^2)$ can be understood more rigorously from the helical geometry shown in Figure \ref{fig:torsion}(b). The metric associated with this structure (Eq. \ref{eq:metric}) establishes an intrinsic coupling between the longitudinal motion along the $z$ axis, characterized by the wave number $k_z$, and the angular motion described by $d \varphi$. This coupling is purely geometric and reflects the fact that the surface torsion induces a correlation between the system's linear and rotational degrees of freedom.

From a physical point of view, when the particle has non-zero momentum along the $z$-axis ( $k_z \neq$ 0 ), the torsion $\tau$ imposes an effective rotation along the azimuthal coordinate, causing the longitudinal motion to be accompanied by an angular displacement. As the particle moves away from the axis, increasing the radial coordinate $\rho$, the kinetic energy associated with this forced rotation increases, resulting in a position-dependent energy term.

\begin{figure}[tbhp]
\centering
\includegraphics[width=\linewidth]{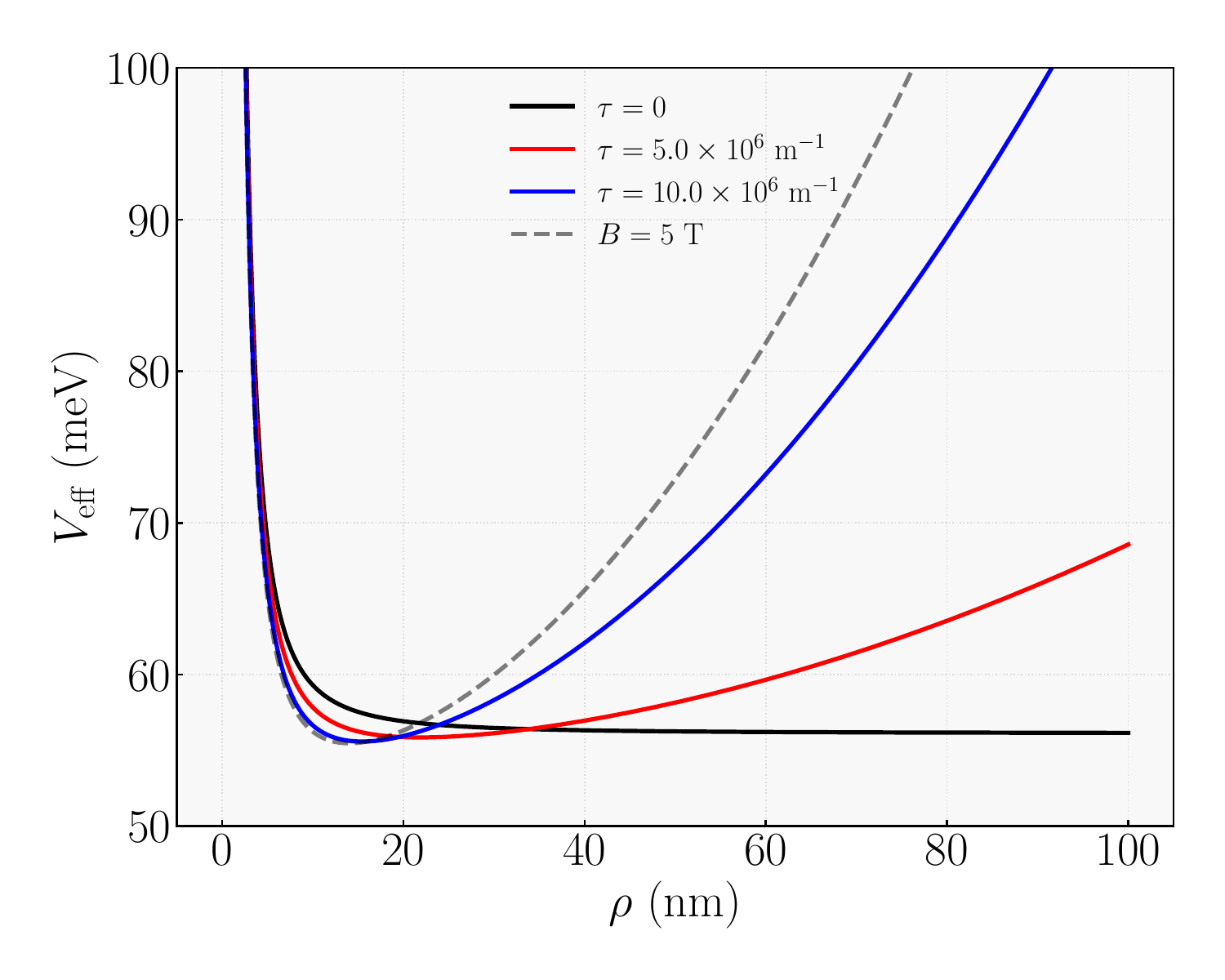}
\caption{Radial effective potential $V_{\mathrm{eff}}(\rho)$ for $m = +1$ and different values of the torsion parameter $\tau$. Colored curves correspond to the case without a magnetic field ($B = 0$), whereas the semi-transparent black dashed curve represents $B = 5\,\mathrm{T}$, illustrating the influence of an external magnetic field on the effective potential.}
\label{fig:Veff}
\end{figure}

This term acts as an effective confinement potential, whose intensity increases with $\rho$, reproducing the characteristic behavior of a harmonic potential. Consequently, the confinement term in the effective potential (Eq. \ref{eq:Veff_pureE}) is proportional to $\Lambda^2$, highlighting its direct dependence on both the geometric torsion $\tau$ and the longitudinal momentum $k_z$. The absence of either factor eliminates coupling, thereby suppressing the geometrically induced confinement. Figure \ref{fig:Veff} illustrates this effect: the increase in $\tau$ intensifies the geometric potential well, reinforcing the particle's confinement.

Figure~\ref{fig:Veff} shows the radial effective potential $(V_{\mathrm{eff}}(\rho))$ for $m = 1$ and different values of the torsion parameter $(\tau)$. Increasing $(\tau)$ significantly alters the potential profile, shifting and reshaping the minima associated with bound states. This behavior reflects the role of torsion in modifying the electronic confinement conditions, a direct consequence of the system's non-Euclidean geometry. The semi-transparent black-dashed curve corresponds to the case with a magnetic field $(B = 5\,\mathrm{T})$, which exhibits an additional shift in the potential due to magnetic coupling. These results highlight the competing influence of torsion and magnetic field on the system’s energy structure, indicating that both parameters can be used to tune the confinement strength and the spatial distribution of electronic states.

Although Eq.~(\ref{eq:Veff_pureE}) provides the effective potential, solving the original radial equation Eq.~(\ref{eq:radial_equation_initial}) directly offers a more standard path to identifying the confluent hypergeometric (Kummer) equation.
For this purpose, we begin by introducing the dimensionless variable $\xi = |\Lambda|\rho^2$. The differential operators with respect to $\rho$ must be transformed into operators with respect to $\xi$. Using the chain rule, the first derivative becomes
\begin{equation}
 \frac{d}{d\rho} = \frac{d\xi}{d\rho}\frac{d}{d\xi} = 2|\Lambda|\,\rho\,\frac{d}{d\xi}.
\end{equation}
The radial part of the Laplacian, $\frac{1}{\rho}\frac{d}{d\rho}(\rho\,dR/d\rho) = d^2R/d\rho^2 + (1/\rho)\,dR/d\rho$, can then be expressed in terms of $\xi$ as
\begin{equation}
 \frac{1}{\rho}\frac{d}{d\rho}\left(\rho\frac{d}{d\rho}\right) = 4|\Lambda|\,\xi\,\frac{d^2}{d\xi^2} + 4|\Lambda|\,\frac{d}{d\xi}.
\end{equation}
Substituting this transformation and the relations $\rho^2 = \xi/|\Lambda|$ and $1/\rho^2 = |\Lambda|/\xi$ into Eq.~\eqref{eq:radial_equation_initial}, and grouping the terms, we arrive at the transformed radial equation
\begin{equation}
 \xi\frac{d^2R}{d\xi^2} + \frac{dR}{d\xi}
 +\left[ -\frac{(m-l)^2}{4\xi} - \frac{\xi}{4} + \beta' \right] R(\xi) = 0,
 \label{eq:transformed_radial_eq}
\end{equation}
where we have defined a new energy-related parameter
\begin{equation}
 \beta' \equiv \frac{1}{4|\Lambda|}\left(\frac{2m^* E}{\hbar^2} - k_z^2 + 2(m-l)\Lambda\right).
\end{equation}

To solve Eq.~\eqref{eq:transformed_radial_eq}, we analyze its asymptotic behavior. For $\xi \to \infty$, the equation approximates to $R'' - (1/4)R \approx 0$, which has a decaying solution $R \sim e^{-\xi/2}$. For $\xi \to 0$, the equation approximates to $\xi R'' + R' - [(m-l)^2/4\xi]\,R \approx 0$, with a regular solution $R \sim \xi^{|m-l|/2}$. Based on this, we propose a solution of the form
\begin{equation}
 R(\xi) = \xi^{|m-l|/2} e^{-\xi/2} u(\xi),
 \label{eq:solution_ansatz}
\end{equation}
where $u(\xi)$ is a function that must be regular at the origin and behave polynomially at infinity for the wave function to be normalizable. Substituting this form back into Eq.~\eqref{eq:transformed_radial_eq} yields the following differential equation for $u(\xi)$:
\begin{equation}
 \xi\,u'' + \left( |m-l| + 1 - \xi \right) u' + \left( \beta' - \frac{|m-l|+1}{2} \right) u = 0,
 \label{eq:kummer_equation}
\end{equation}
which is the standard form of the confluent hypergeometric (Kummer) equation. Its general solution is the confluent hypergeometric function $F(a,b,\xi)$, with parameters $a = (|m-l|+1)/2 - \beta'$ and $b = |m-l|+1$.

The physically acceptable solutions, finite at the origin and vanishing at infinity, require the series expansion of the confluent hypergeometric function to terminate, which occurs when its first parameter $a$ is a non-positive integer. We therefore impose the quantization condition
\begin{equation}
 a = -n, \qquad n=0,1,2,\dots
\end{equation}
which restricts the possible values of $\beta'$ and quantizes the system's energy levels $E$. When this condition is met, $u(\xi)$ becomes proportional to the generalized Laguerre polynomial, $L_n^{p}(\xi)$, through the identity $L_n^{p}(\xi) \propto F(-n, p+1, \xi)$, with $p=b-1=|m-l|$.

Thus, the radial part of the wave function is described by Laguerre polynomials. By applying the normalization condition $\int_0^\infty |R(\rho)|^2 \rho\, d\rho = 1$, we obtain the normalized radial wave functions (see \ref{app:normalization})
\begin{align}
 R_{n,m}(\rho) =
 \sqrt{\frac{2|\Lambda|\, n!}{\Gamma\big(n+|m-l|+1\big)}}\;
 &\big(|\Lambda|\rho^2\big)^{\frac{|m-l|}{2}}
 \exp\left(-\frac{|\Lambda|\rho^2}{2}\right)
 \notag\\ &\times L_n^{(|m-l|)}\big(|\Lambda|\rho^2\big).
 \label{eq:wavefunction_explicit}
\end{align}
This expression will be used to compute the probability amplitude for finding the electron at a radial distance $\rho$ from the center of the topological defect.

The corresponding quantized energy eigenvalues are found to be
\begin{align}
 E_{n,m}
 &= \left( \frac{\hbar^2 k_z \tau}{m^{*}} + \frac{\hbar e B}{2m^{*}} \right)
 \left( 2n + |m-l| - (m-l) + 1 \right)
 + \frac{\hbar^2 k_z^2}{2m^*},
\end{align}
or equivalently
\begin{align}
 E_{n,m}
 &= \hbar\left(\omega_{\tau}+\omega_{c}\right)
 \left(n+\frac{|m-l|}{2}-\frac{(m-l)}{2}+\frac{1}{2}\right)
 + \frac{\hbar^{2}k_{z}^{2}}{2m^{*}},
 \label{eq:energy_eigenvalues}
\end{align}
where $\omega_c = eB/m^{*}$ is the cyclotron frequency and
\begin{equation}
 \omega_\tau \equiv \frac{\hbar k_z \tau}{m^*}.
\end{equation}
This expression reveals that the energy spectrum is directly influenced by the interplay between the material's torsion ($\tau$), the external magnetic field ($B$), and the AB flux parameter $l$.
\begin{figure}[tbhp]
\centering
\includegraphics[width=0.45\textwidth]{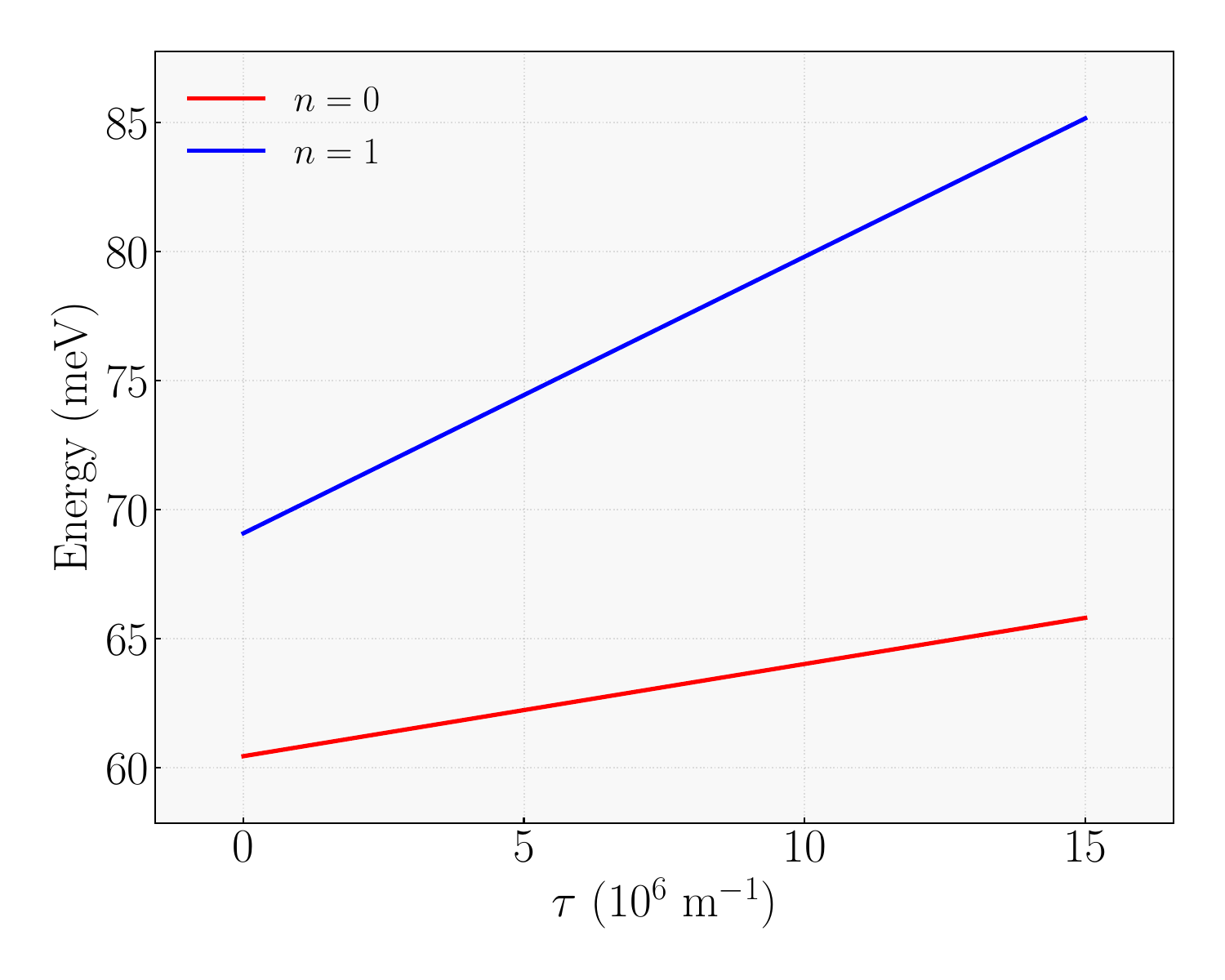}
\caption{Energy eigenvalues for the ground state ($n=0$, red) and the first radially excited state ($n=1$, blue) as a function of the torsion density $\tau$. The calculations are performed for a fixed azimuthal quantum number $m=0$, Aharonov--Bohm flux $l=0.1$ ($h/e$), and magnetic field $B=5$~T. The plot shows that the energy of both levels increases with torsion. The energy gap between the two states also widens, indicating a stronger torsional influence on the excited state. The linear increase in energy demonstrates the role of torsion as an effective confining potential.}
\label{fig:energy_levels}
\end{figure}
\begin{figure*}[tbhp]
    \centering
    \includegraphics[width=1\linewidth]{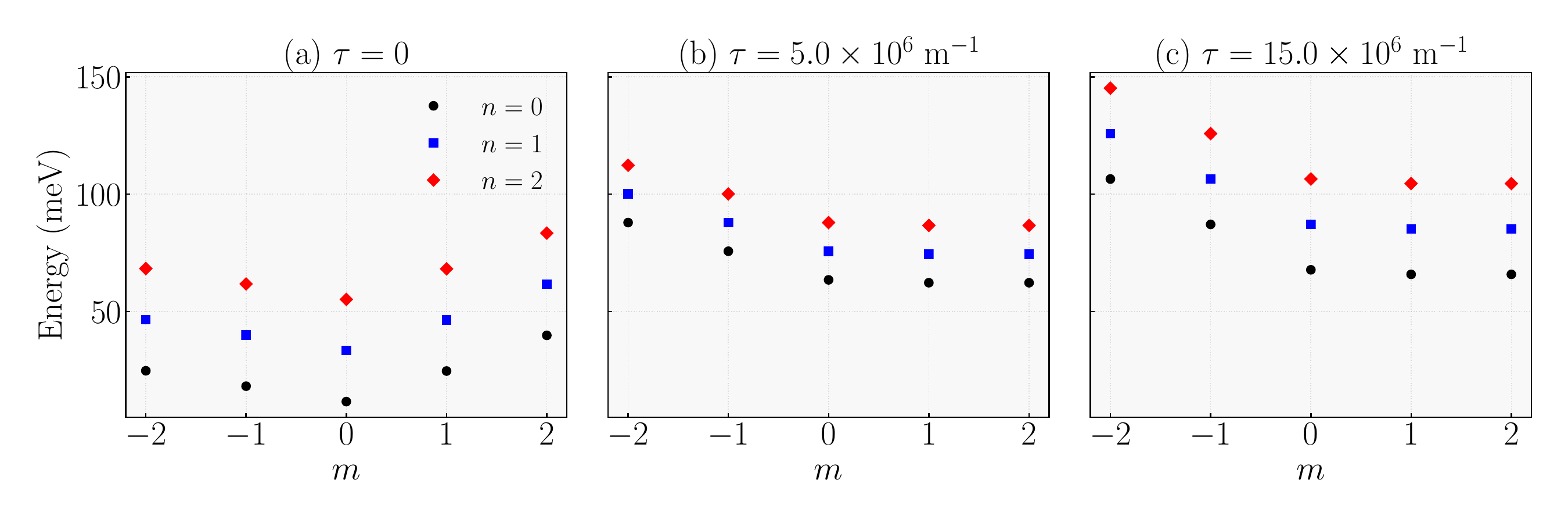}
    \caption{Energy eigenvalues as a function of the magnetic quantum number $m$ ($m=-2,-1,0,1,2$), for three different values of torsion density: (a) $\tau = 0$, (b) $\tau = 5.0 \times 10^6\,\mathrm{m}^{-1}$ and (c) $\tau = 15.0 \times 10^6\,\mathrm{m}^{-1}$. It can be seen that the energy values increase with increasing $\tau$.}
    \label{fig:energy_m}
\end{figure*}

The dependence of the energy eigenvalues on the torsion density $\tau$ is presented in Fig.~\ref{fig:energy_levels}. The plot shows the energies of the ground state ($n=0$) and the first radially excited state ($n=1$), both for an electron with azimuthal quantum number $m=0$, a fixed AB flux of $l=0.1$ ($h/e$), and an external magnetic field of $B=5.0$~T.

The most prominent feature is the significant increase in the energy for both the ground state and the first excited state as the torsion density $\tau$ grows. This behavior is a direct consequence of torsion serving as an effective confining potential. A larger value of $\tau$ corresponds to a stronger spatial confinement, which compresses the electron's wave function into a smaller region. In accordance with the principles of quantum mechanics, localizing a particle more strongly increases its kinetic energy, thereby raising its total energy eigenvalue. For the range of parameters considered, this increase is approximately linear with $\tau$.

Furthermore, the graph reveals that the energy difference between states $n=1$ and $n=0$, $\Delta E = E_1 - E_0$, is not constant but increases with increasing twist. This indicates that twist affects the excited state more strongly than the ground state, further separating the energy levels.

Complementarily, Fig.~\ref{fig:energy_m} shows the energy eigenvalues as a function of the magnetic quantum number $m$, a crucial parameter for the optical transition of the system according to Fermi's golden rule. The graph was obtained for three different values of torsion density: (a) $\tau = 0$, (b) $\tau = 5.0 \times 10^6\,\mathrm{m}^{-1}$ and (c) $\tau = 15.0 \times 10^6\,\mathrm{m}^{-1}$. It can be observed that the energy values increase with the increment of $\tau$, revealing differences between states with the same value of $n$. For visual distinction, circles, squares, and diamonds represent, respectively, the quantum numbers $n=0$, $n=1$, and $n=2$.

This tunable energy gap has direct implications for the system's optical properties. As observed in the photoionization calculations, the resonance peak blueshifts with increasing $\tau$, a direct consequence of the increasing transition energy.
\begin{figure}[tbhp]
 \centering
 \includegraphics[width=0.48\textwidth]{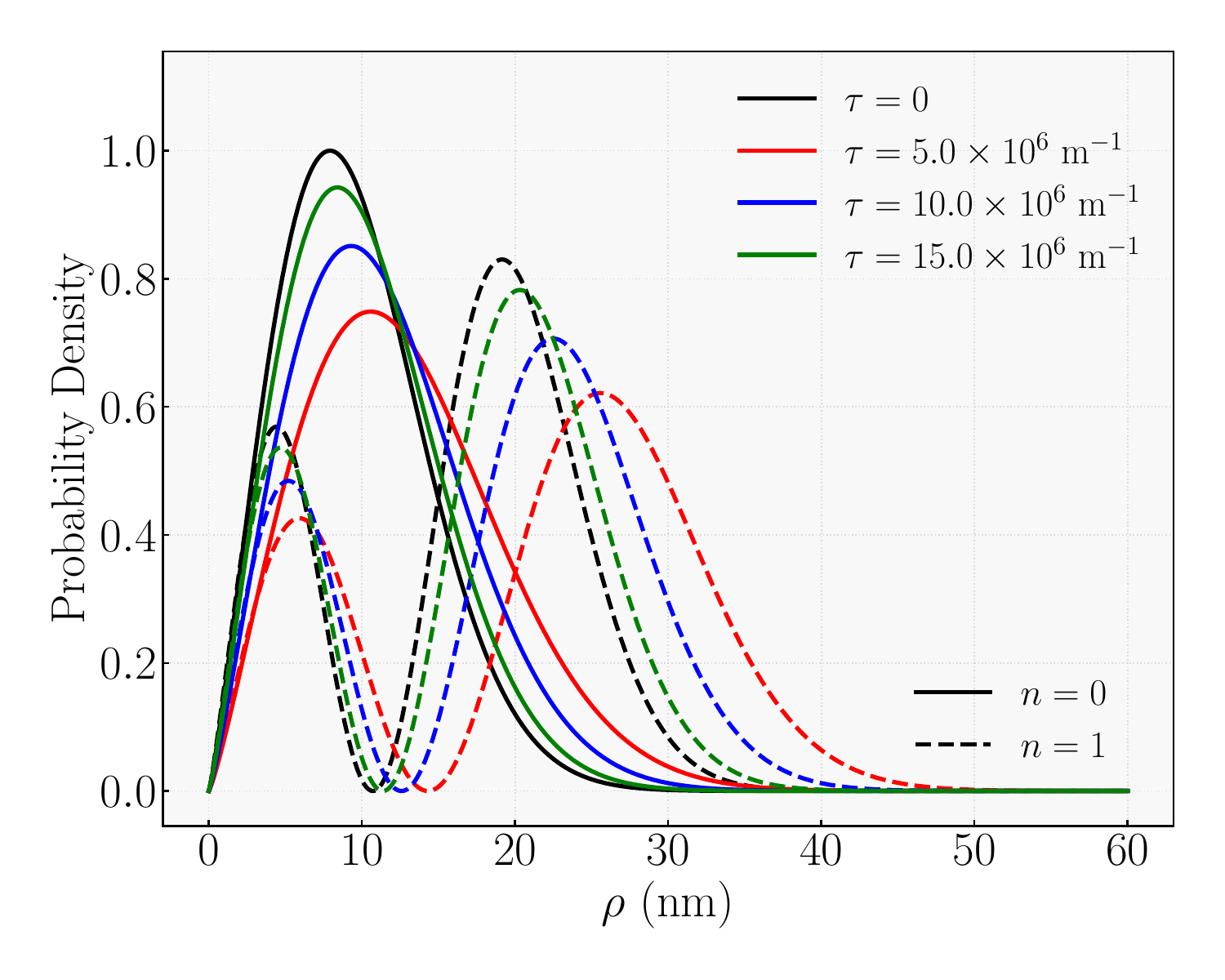}
 \caption{Normalized radial probability density for the $n=0$ (solid lines) and $n=1$ (dashed lines) states, calculated for a fixed flux $l=0.1$ ($h/e$) and magnetic field $B=5$~T. The colors correspond to different values of the torsion density $\tau$. All curves have been normalized by the peak value of the torsion-free ground state ($n=0$, $\tau=0$, solid black line). The plot highlights the suppression of the probability density due to both radial excitation (lower peaks for $n=1$) and increasing torsion (lower peaks for colored curves).}
 \label{fig:prob_density_global_norm}
\end{figure}

The effect of the torsion on the electron's spatial distribution is analyzed by examining the radial probability density, $P_{n,m}(\rho) = \rho\,|R_{n,m}(\rho)|^2$. Figure~\ref{fig:prob_density_global_norm} displays this quantity for the ground state ($n=0$) and the first radially excited state ($n=1$), both with $m=0$, for a fixed AB flux of $l=0.1$ ($h/e$) and a magnetic field of $B=5$~T.

To provide a direct comparison of all states relative to the absolute ground state of the torsion-free system, a global normalization has been applied. All curves shown in the plot have been divided by the peak value of the probability density for the $\tau=0$, $n=0$ case (the solid black line). Consequently, this reference curve has a peak of exactly unity, and all other state probabilities are shown as a fraction of this absolute maximum.

This visualization clearly separates the effects of radial excitation from those of torsion. The impact of radial excitation from $n=0$ (solid lines) to $n=1$ (dashed lines) is immediately apparent. The $n=1$ state, which possesses one radial node, has a significantly lower probability density peak and is distributed over a larger radial distance. This illustrates the spreading of the wave function for higher energy levels.

The influence of the torsion density $\tau$ manifests as both a spatial confinement and a strong amplitude suppression. As $\tau$ increases (from black to the colored curves), the peaks for both the $n=0$ and $n=1$ states shift to smaller radial distances, confirming the role of torsion as a confining potential. Concurrently, the peak heights are dramatically reduced relative to the global maximum. This strong suppression of the probability density provides a clear visual explanation for the weakening of the optical transition matrix elements that was observed in the photoionization and absorption calculations.

With the system's eigenstates and energy levels now defined, we are in a position to calculate how it responds to electromagnetic radiation, which will be the focus of the next section.

\section{Optical Properties: Theoretical Framework}
\label{sec:optical_properties_framework}

The optical response of a confined quantum system, such as a quantum dot or ring, subjected to torsion, magnetic field, and AB flux, can be significantly modified due to changes in the electronic structure and transition matrix elements. In this section, we present the theoretical framework for describing the linear and third-order nonlinear optical absorption coefficients (OACs), the refractive-index changes (RICs), and the photoionization cross-section (PCS).

\subsection{Linear and Nonlinear Optical Absorption and Refraction}
\label{sec:optical_properties}

We follow the formalism developed in Refs.~\cite{APA.2023.129.188,PB.2025.716.417733,CTP.2024.76.105701,QR.2024.6.677,PLA.2023.466.128725,PM.2021.101.2614,PE.2018.95.27} and assume the system interacts with an external monochromatic electric field of the form:
\begin{equation}
 \boldsymbol{E}(t) = E_0 \hat{e}_r \cos(\omega t),
\end{equation}
where $E_0$ is the field amplitude, $\omega$ is the angular frequency of the radiation, and the polarization is taken in the radial direction $\hat{e}_r$.

The interaction Hamiltonian is treated semi-classically, and the transition rate between quantum states $\ket{\psi_i}$ and $\ket{\psi_f}$ is obtained via Fermi’s golden rule, assuming electric dipole transitions. The interaction induces transitions governed by the dipole matrix element:
\begin{equation}
 M_{if} = \bra{\psi_f} \,\hat{\boldsymbol r}\, \ket{\psi_i},
\end{equation}
which carries the selection rule $\Delta m = \pm 1$ \cite{APA.2023.129.188,PB.2025.716.417733,CTP.2024.76.105701} for circular or radial polarization, in accordance with the system's cylindrical symmetry.

The total absorption coefficient $\alpha(\omega, I_0)$, including linear and third-order nonlinear contributions, is given by:
\begin{equation}
 \alpha(\omega, I_0) = \alpha^{(1)}(\omega) + \alpha^{(3)}(\omega, I_0),
\end{equation}
where $I_0$ is the intensity of the incident electromagnetic field.

The linear absorption coefficient is:
\begin{equation}
 \alpha^{(1)}(\omega) = \omega \sqrt{\frac{\mu}{\epsilon_r}} \frac{\sigma_s |M_{if}|^2 \hbar \Gamma}{(E_f - E_i - \hbar \omega)^2 + (\hbar \Gamma)^2},
\end{equation}
where $\mu$ is the magnetic permeability, $\epsilon_r$ is the relative permittivity, $n_r$ is the refractive index of the medium, $c$ is the speed of light, $\Gamma$ is the phenomenological broadening parameter (linked to the relaxation time), and $\sigma_s$ is the surface electron density.

The third-order nonlinear contribution arises from the saturation of the medium and multiphoton effects, and is given by
\begin{align} \alpha^{(3)}(\omega, I_0) &= - \omega \sqrt{\frac{\mu}{\epsilon_r}} \left( \frac{I_0}{2 n_r \epsilon_0 c} \right) \sigma_s \frac{|M_{if}|^2 \hbar \Gamma}{\left[(E_f - E_i - \hbar\omega)^2 + (\hbar \Gamma)^2\right]^2} \nonumber \\ &\quad \times \left[ 4|M_{if}|^2 - |M_{ff} - M_{ii}|^2 \left(\cdots \right)\right]. \end{align} 
where 
\begin{equation}
\left(\cdots \right) = \left( \frac{3(E_f - E_i)^2 - 4(E_f - E_i)\hbar \omega + \hbar^2(\omega^2 - \Gamma^2)}{(E_f - E_i)^2 + (\hbar \Gamma)^2} \right).    
\end{equation}

This expression accounts for intensity-dependent phenomena such as saturation, bleaching, and virtual transitions. The relative refractive-index change is derived from the real part of the third-order susceptibility $\chi^{(3)}(\omega)$ and is written as:
\begin{equation}
 \frac{\Delta n(\omega)}{n_r} = \frac{\Delta n^{(1)}(\omega)}{n_r} + \frac{\Delta n^{(3)}(\omega, I_0)}{n_r}.
\end{equation}

The linear part is:
\begin{equation}
 \frac{\Delta n^{(1)}(\omega)}{n_r} = \frac{\sigma_s |M_{if}|^2}{2 n_r^2 \epsilon_0}
 \left( \frac{E_f - E_i - \hbar\omega}{(E_f - E_i - \hbar\omega)^2 + (\hbar \Gamma)^2} \right),
\end{equation}
while the third-order term is:
\begin{align}
 &\frac{\Delta n^{(3)}(\omega, I_0)}{n_r} = -\frac{\mu c \sigma_s I_0}{4 n_r^3 \epsilon_0} \frac{|M_{if}|^2}{\left[ (E_f - E_i - \hbar\omega)^2 + (\hbar \Gamma)^2 \right]^2} \nonumber \\
 &\times \left[ 4(E_f - E_i - \hbar\omega) |M_{if}|^2 - |M_{ff} - M_{ii}|^2 G(\omega) \right],
\end{align}
where the function $G(\omega)$ includes higher-order contributions of the energy detuning and damping.

\subsection{Theory of the Photoionization Cross-Section}
\label{sec:theory_photoionization}

The photoionization process is an optical transition in which an electron, initially in a bound ground state, is excited to a higher-energy state by absorbing a photon. The photoionization cross-section, $\sigma$, quantifies the probability of this process.

Starting from Fermi's golden rule and applying the dipole approximation, the general expression for the photoionization cross-section, $\sigma(\hbar\omega)$, as a function of the photon energy $\hbar\omega$ is given by \cite{CTP.2024.76.105701}
\begin{align}
\sigma(\hbar\omega) = \left[ \left( \frac{F_{\text{eff}}}{F_0} \right)^2 \frac{n_r}{\kappa} \right]& \frac{4\pi^2}{3} \beta_{FS} \hbar\omega \notag\\ & \times\sum_f \abs{ \bra{\psi_i} \boldsymbol{r} \ket{\psi_f} }^2 \delta(E_f - E_i - \hbar\omega).
\label{eq:photoionization_general}
\end{align}
Here, $n_r$ is the refractive index, $\beta_{FS} = e^2/\hbar c$ is the fine-structure constant, and $F_{\text{eff}}/F_0$ is the ratio of the effective electric field to the average field. The core of the expression involves the squared dipole matrix element $\abs{\bra{\psi_i} \boldsymbol{r} \ket{\psi_f}}^2$ between the initial state $\ket{\psi_i}$ and final state $\ket{\psi_f}$, and the Dirac delta function, which ensures energy conservation.

For numerical calculations, the local field ratio $F_{\text{eff}}/F_0$ is approximated as unity. A crucial step is the replacement of the Dirac delta function with a narrow Lorentzian profile to account for the finite lifetime of the excited state:
\begin{equation}
\delta(E_f - E_i - \hbar\omega) \rightarrow \frac{1}{\pi} \frac{\hbar\Gamma}{(E_f - E_i - \hbar\omega)^2 + (\hbar\Gamma)^2},
\label{eq:lorentzian}
\end{equation}
where $\hbar\Gamma$ is a phenomenological broadening parameter.

In this work, we investigate the transition from the ground state ($n=0, m=0$) to the first optically accessible excited state ($n=0, m=-1$), according to the dipole selection rule $\Delta m = \pm 1$. The final expression used for the calculations is
\begin{align}
\sigma(\hbar\omega) = \frac{n_r}{\kappa} \frac{4\pi}{3} \beta_{FS} \hbar\omega &\abs{\bra{\Psi_{0,-1}} \boldsymbol{r} \ket{\Psi_{0,0}}}^2 \notag\\ & \times \frac{\hbar\Gamma}{(E_{0,-1} - E_{0,0} - \hbar\omega)^2 + (\hbar\Gamma)^2},
\label{eq:photoionization_final}
\end{align}
where the initial and final states are denoted by their respective $(n,m)$ quantum numbers.

\section{Results and Discussion}
\label{sec:results}

In this section, we present and discuss the numerical results for the optical properties of the system, including the nonlinear absorption coefficients, refractive index changes, and the photoionization cross-section. These results are particularly relevant for quantum dot systems based on materials like InAs/GaAs, where structural defects and external fields can be experimentally realized.
Before presenting the numerical results, it is pertinent to emphasize the essential role played by the wave number $(k_{z})$ in the description of the system. According to the theoretical model developed, the effect of torsion $(\tau)$ is intrinsically associated with the longitudinal movement of the electron along the $(z)$-axis. This coupling manifests explicitly in the characteristic factor $\Lambda$, as well as in the torsion-dependent effective frequency $\omega_{\tau}$.

The presence of this term reveals a central physical aspect of the model: torsion only influences the energy spectrum when there is a finite momentum component along the $(z)$-axis. In other words, for $(k_{z}=0)$, the energy spectrum (see Eq. \ref{eq:energy_eigenvalues}) becomes independent of the torsion parameter, indicating that the geometric confinement induced by torsion is a phenomenon that emerges exclusively in states with longitudinal propagation. This behavior is consistent with previous results on geometric couplings in confined quantum systems, in which the medium's curvature and torsion produce effective potentials that depend on the longitudinal coordinate.

Therefore, in all subsequent simulations, we adopted a finite and representative value for the wave number, typically $(k_{z}=10^{9}\pi~\mathrm{m^{-1}})$, to ensure that the effects of torsion on the optical properties are properly captured and quantitatively analyzed.
\begin{figure*}[tbhp]
\centering
\includegraphics[width=0.48\textwidth]{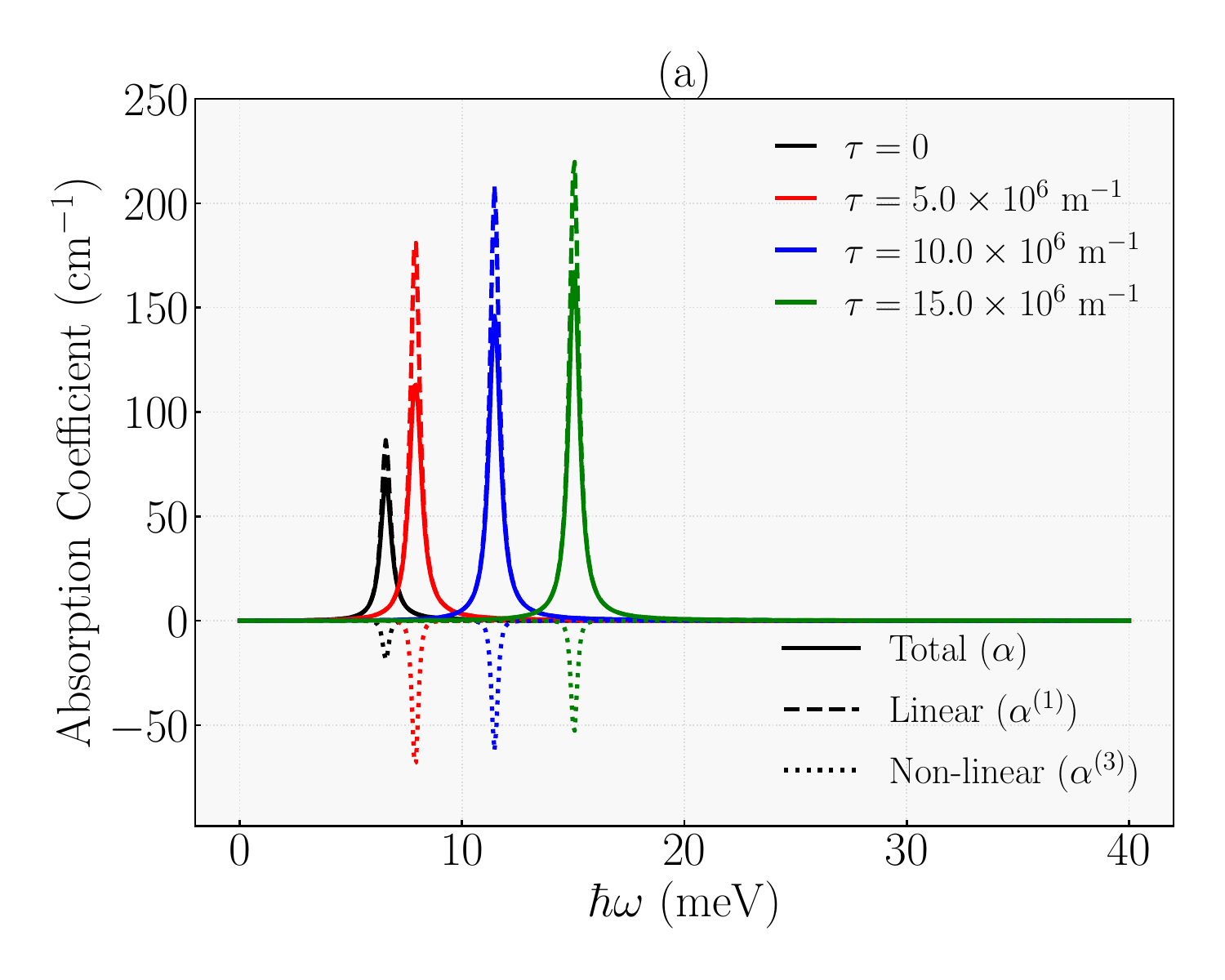}
\includegraphics[width=0.48\textwidth]{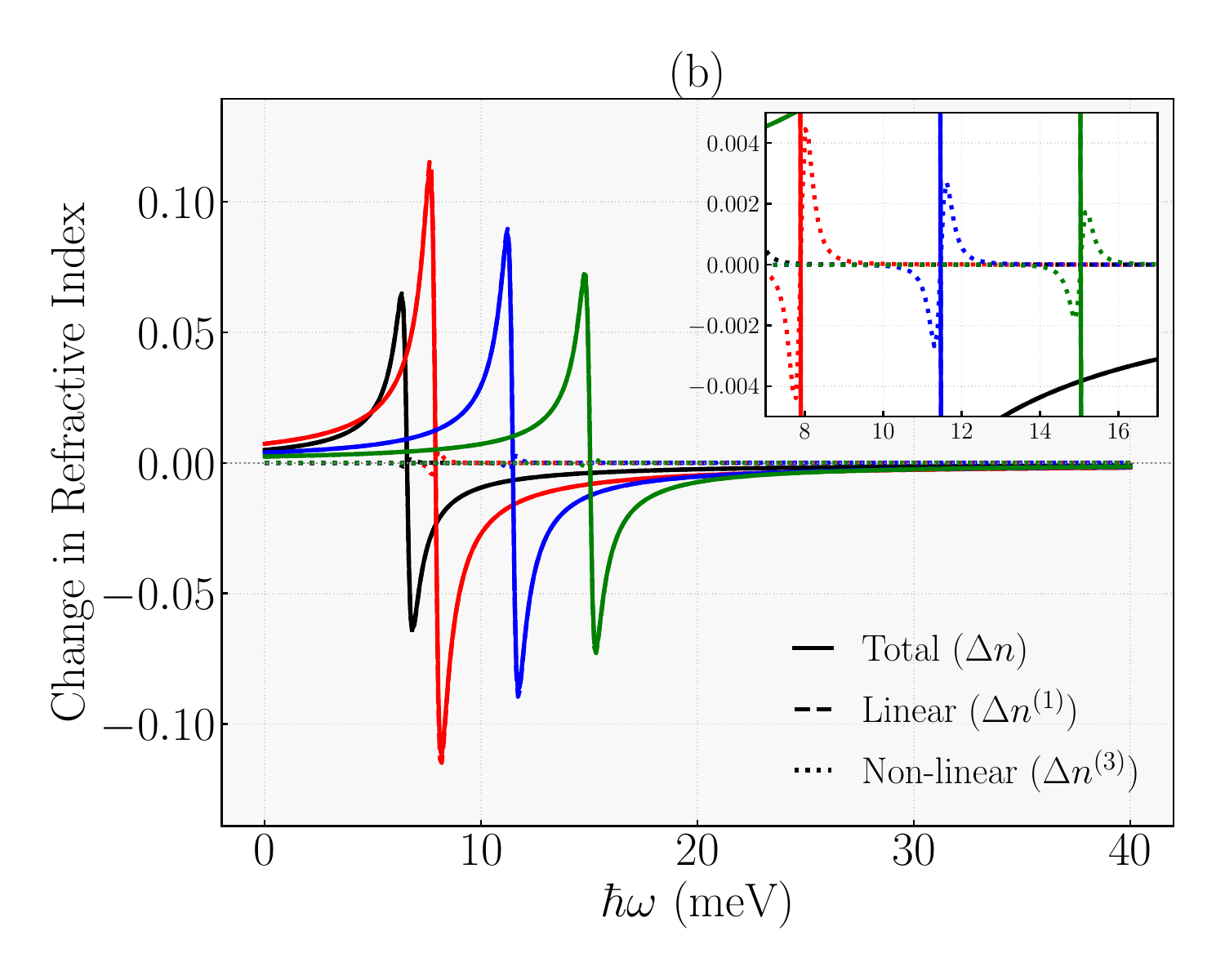}
\caption{Nonlinear optical properties as a function of photon energy. The curves represent different values of the torsion density $\tau$. In Fig. (a), we plot the optical absorption coefficient and, in Fig. (b), the total refractive index change. In both figures, a shift of the peaks to higher energies (blueshift) and a suppression of the transition amplitude are observed with increasing $\tau$. We use $I_0 = 5.0 \times 10^5$ W/m$^2$, $l = 0.1$, $\hbar \omega_{0}=10$ meV and $B=5.0$~T.}
\label{fig:optics}
\end{figure*}

The nonlinear optical properties are presented in Fig. \ref{fig:optics}. The calculations were performed for a system subjected to a magnetic field of $B = 5$~T and a fixed AB flux of $l = 0.1$, considering the transition from the ground state ($n=0, m=0$) to the first excited state ($n=0, m=-1$). The different curves in each panel correspond to different values of the medium's torsion density, $\tau$.

\begin{figure}[tbhp]
\centering
\includegraphics[width=0.48\textwidth]{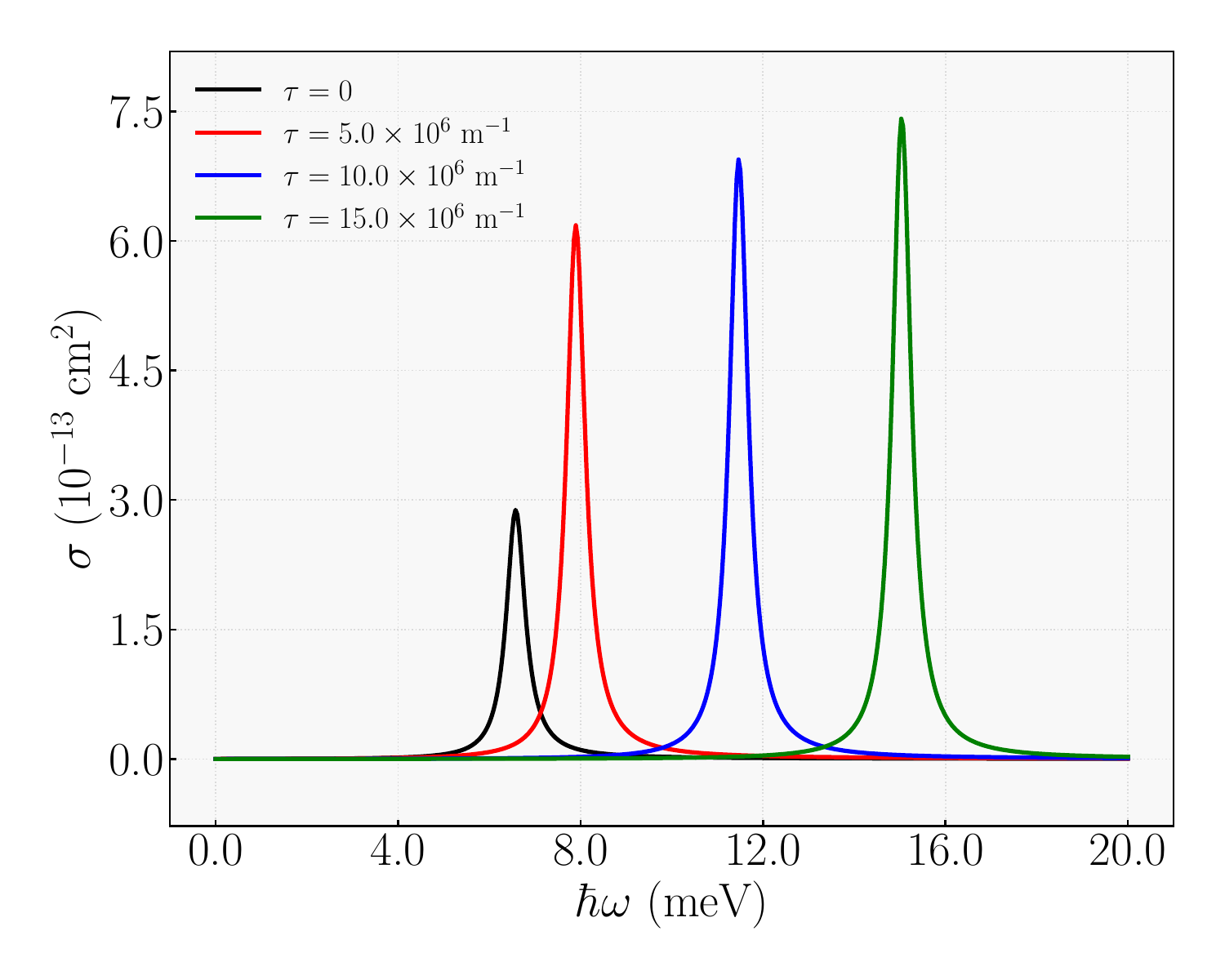}
\caption{Photoionization cross-section ($\sigma$) as a function of the incident photon energy ($\hbar\omega$). The calculations were performed for a fixed Aharonov--Bohm flux of $l = 0.1$, a magnetic field of $B=5$~T, $\hbar \omega_{0}= 10$ meV, and $k_z = 10^{9}\pi\, \mathrm{m}^{-1}$. The different curves correspond to varying values of the torsion density $\tau$, as indicated in the legend. The plot shows both a blueshift of the resonance peak and a suppression of its amplitude as the torsion density increases.}
\label{fig:photoionization}
\end{figure}

Figure \ref{fig:optics}(a) displays the optical absorption coefficients. The black curve represents the reference case without torsion ($\tau = 0$), where a pronounced resonance peak is observed at approximately 15~meV. Upon introducing and increasing the torsion density $\tau$ (red, blue, and green curves), two main features stand out. First, a blueshift is observed: the position of the resonance peak shifts to higher energies as $\tau$ increases. This behavior is a direct reflection of the increase in the transition energy, $\Delta E = E_1 - E_0$. Second, a significant amplitude suppression occurs. The intensity of the absorption peak is drastically reduced with increasing $\tau$. This effect is attributed to a decrease in the transition dipole matrix element ($M_{21}$), as torsion alters the spatial shapes of the wave functions and diminishes their overlap. This same blueshift and amplitude suppression behavior is confirmed directly in the PCS calculations, as shown in Fig.~\ref{fig:photoionization}. Both calculations, therefore, consistently demonstrate how the $\tau$ torsion affects the $\Delta m = -1$ transition.

Figure \ref{fig:optics}(b) shows the refractive index change. The curves exhibit the characteristic dispersive shape associated with resonant absorption. The effects of the torsion $\tau$ are consistent with those observed for the absorption. The entire dispersive structure shifts to higher energies with increasing $\tau$, and the magnitude of the refractive index change is significantly attenuated. Taken together, the results demonstrate that the torsion density $\tau$ acts as an effective control parameter for the system's optical properties.
\begin{figure*}[tbhp]
\centering
\includegraphics[width=0.48\textwidth]{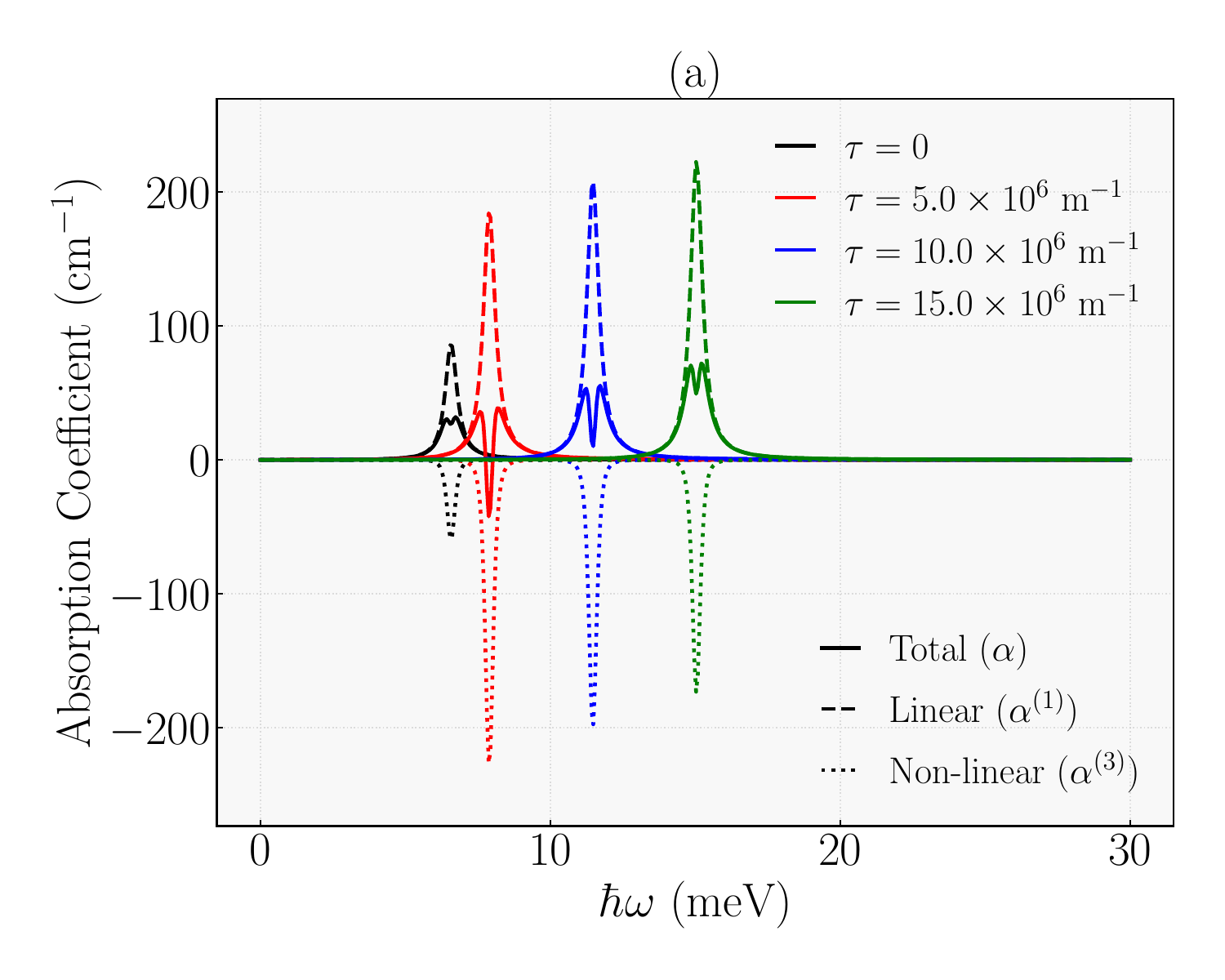}
\includegraphics[width=0.48\textwidth]{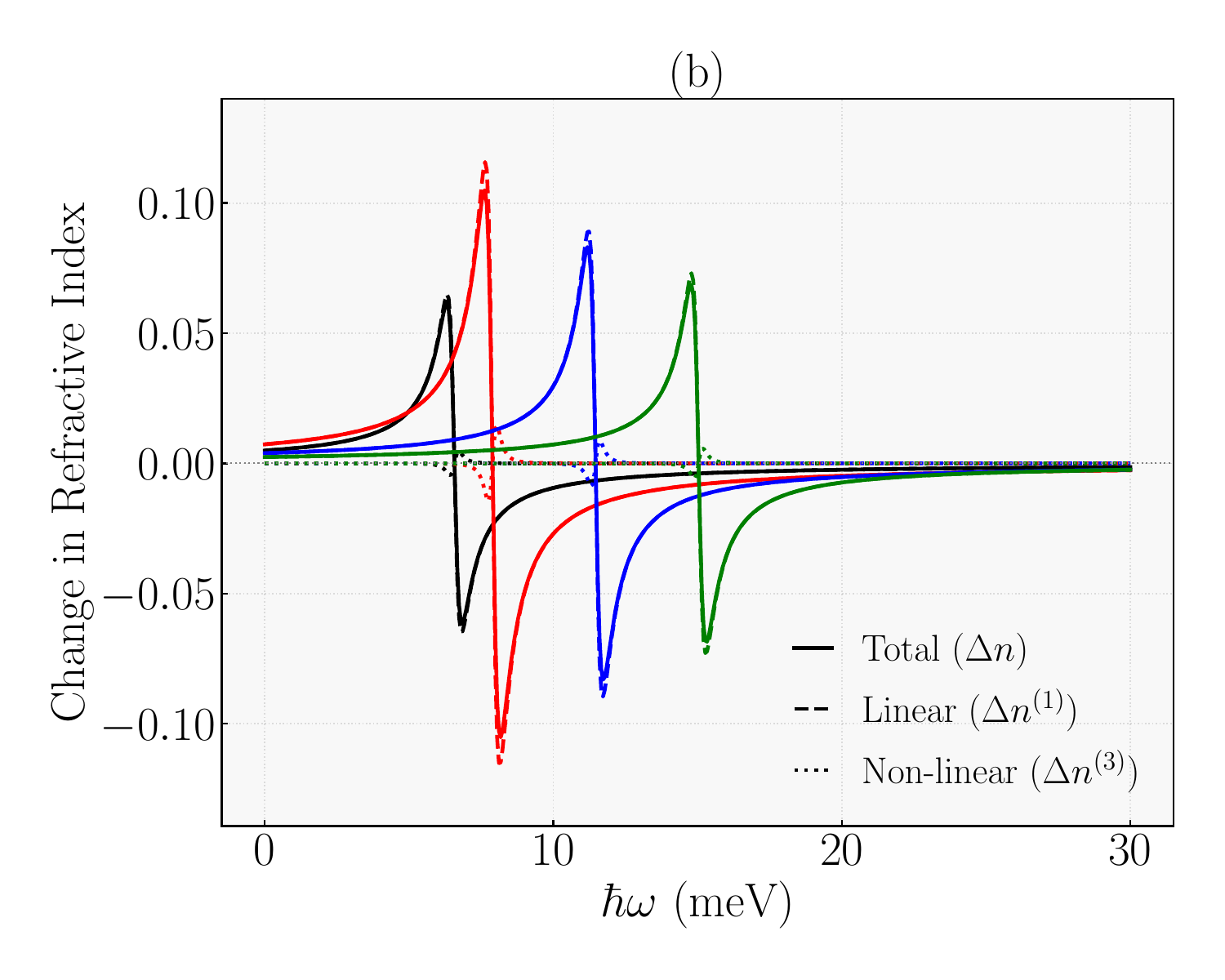}
\caption{Optical properties in the high-intensity regime. (a) Absorption coefficient, exhibiting the phenomenon of optical switching (negative absorption or gain) at the center of the resonance for lower values of $\tau$. (b) Refractive index change, showing an enhanced and sharper dispersive profile due to the strong nonlinear contribution. We use $I_0 = 1.6 \times 10^6$ W/m$^2$, $l = 0.1$, $\hbar \omega_{0}=10$ meV and $B=5$~T.}
\label{fig:optics_high_I0}
\end{figure*}

By increasing the incident light intensity to $I_0 = 1.6 \times 10^6$ W/m$^2$, the system enters a strongly nonlinear regime, as seen in Fig.~\ref{fig:optics_high_I0}. Figure~\ref{fig:optics_high_I0}(a) shows that for lower torsion values (black and red curves), the absorption peak inverts, becoming negative at the resonance center. This phenomenon of optical switching occurs because the nonlinear term $\alpha^{(3)}$, which is negative and proportional to $I_0$, becomes larger in magnitude than the positive linear term $\alpha^{(1)}$, resulting in optical gain ($\alpha < 0$).

Figure~\ref{fig:optics_high_I0}(b) illustrates the refractive index change in this high-intensity regime. The nonlinear contribution $\Delta n^{(3)}$ has a sharper dispersive shape than the linear term $\Delta n^{(1)}$. The superposition of the two terms results in a total $\Delta n$ profile with an enhanced peak-to-valley amplitude and a noticeably sharper, more distorted shape, which is a clear signature of the strong nonlinear contribution.
\begin{figure*}[tbhp]
\centering
\includegraphics[width=0.48\textwidth]{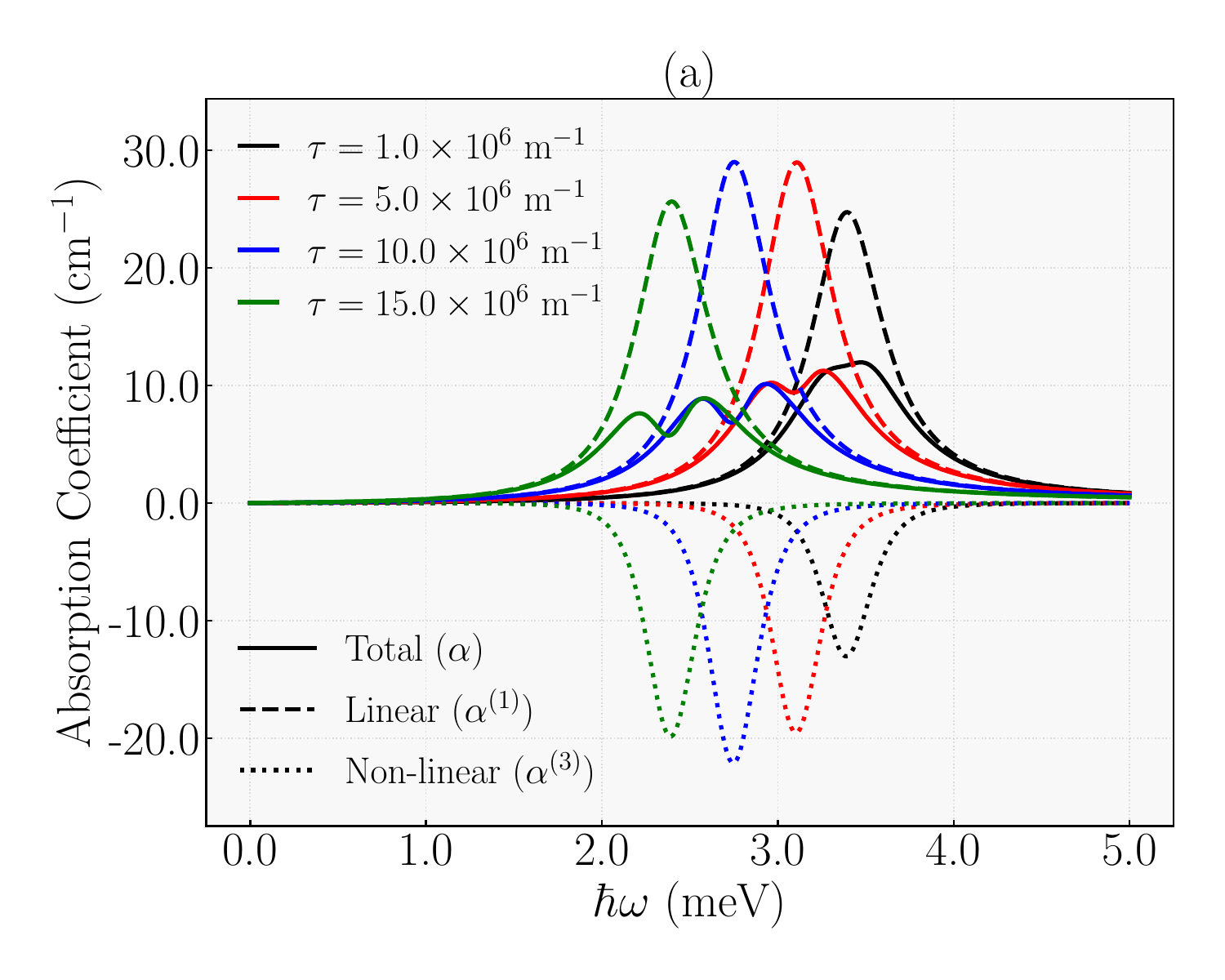}
\includegraphics[width=0.48\textwidth]{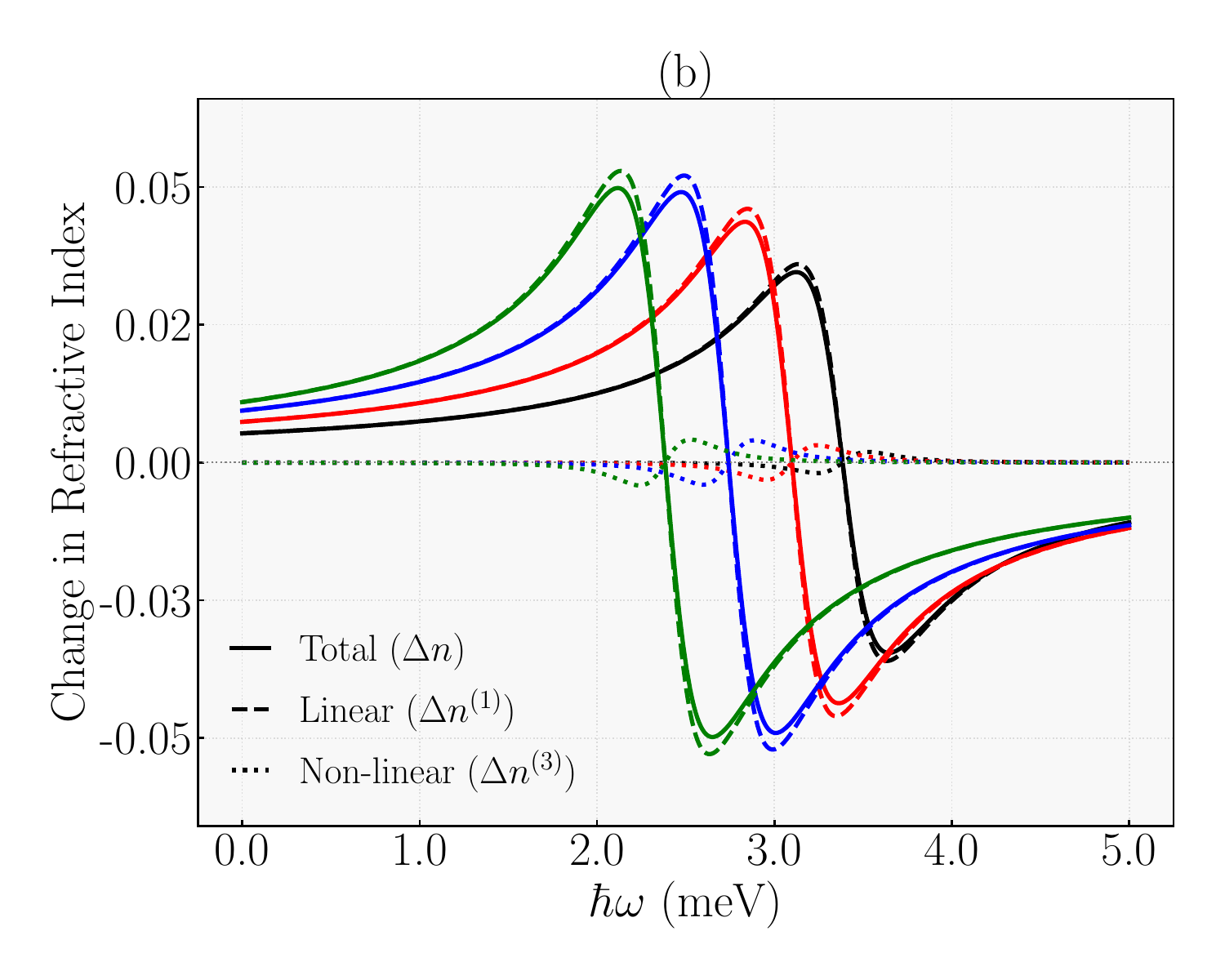} \\
\caption{Optical absorption and refractive index change coefficients for an electron confined in a quantum dot in the presence of a topological defect (torsion). The panels show (a) the absorption coefficient and (b) the change in refractive index as a function of the incident photon energy ($\hbar\omega$) for the $m=1$ angular momentum transition. The curves are calculated for different values of torsion density $\tau$ (indicated in the legend of (a)). We use $I_0 = 22 \times 10^5$ W/m$^2$, $\hbar \omega_{0}=10$ meV and $B=5$~T.}
\label{fig:optical_properties}
\end{figure*}

Figure~\ref{fig:optical_properties} further illustrates the impact of spatial torsion on the nonlinear optical properties for the transition to the $m=1$ excited state. The panels display the absorption coefficient (a) and the change in the refractive index (b) as a function of the incident photon energy, $\hbar\omega$. The linear ($\alpha^{(1)}$, $\Delta n^{(1)}$), third-order nonlinear ($\alpha^{(3)}$, $\Delta n^{(3)}$), and total ($\alpha$, $\Delta n$) contributions are represented by dashed, dotted, and solid lines, respectively. The primary effect observed is that an increase in the torsion density $\tau$ induces a blueshift of the resonance peaks. This is a direct consequence of the elevated transition energy $\Delta E$. Beyond the energy shift, the amplitude of the peaks is also modulated by torsion, reflecting the alteration in the transition dipole matrix element.

\begin{table*}[tbhp]
\centering
\caption{Comparison of the transition energy differences (in meV) in the presence ($\ell = 0.1$) and absence ($\ell = 0$) of the Aharonov--Bohm flux, for selected values of torsion ($\tau$) and magnetic field ($B$), according to the corrected analytical model. Parameters: $n=0$, $k_z=\pi\times10^{9}\,\mathrm{m}^{-1}$, $\mu=0.067\,m_e$.}\vspace{0.5cm}
\label{tab:comparacao_fluxo_completa}
\begin{tabular}{cc|cc|cc}
\hline\hline
\multirow{2}{*}{$\tau\times 10^6 \mathrm{~m}^{-1}$} & \multirow{2}{*}{$B\, (\mathrm{T})$} & \multicolumn{2}{c|}{Flux ($\ell = 0.1$)} & \multicolumn{2}{c}{Flux ($\ell = 0$)} \\
\cline{3-6}
 & & $\Delta E_{0 \to 1}$ & $\Delta E_{0 \to -1}$ & $\Delta E_{0 \to 1}$ & $\Delta E_{0 \to -1}$ \\
\hline
\multirow{4}{*}{$0.00$} & 0.0 & 0.000000 & 0.000000 & 0.000000 & 0.000000 \\
 & 1.0 & -0.172788 & 1.727875 & 0.000000 & 1.727875 \\
 & 2.5 & -0.431969 & 4.319688 & 0.000000 & 4.319688 \\
 & 5.0 & -0.863938 & 8.639376 & 0.000000 & 8.639376 \\
\hline
\multirow{4}{*}{$5.00$} & 0.0 & -3.572959 & 35.729588 & 0.000000 & 35.729588 \\
 & 1.0 & -3.745746 & 37.457463 & 0.000000 & 37.457463 \\
 & 2.5 & -4.004928 & 40.049275 & 0.000000 & 40.049275 \\
 & 5.0 & -4.436896 & 44.368963 & 0.000000 & 44.368963 \\
\hline
\multirow{4}{*}{$10.00$} & 0.0 & -7.145918 & 71.459175 & 0.000000 & 71.459175 \\
 & 1.0 & -7.318705 & 73.187050 & 0.000000 & 73.187050 \\
 & 2.5 & -7.577886 & 75.778863 & 0.000000 & 75.778863 \\
 & 5.0 & -8.009855 & 80.098551 & 0.000000 & 80.098551 \\
\hline
\multirow{4}{*}{$15.00$} & 0.0 & -10.718876 & 107.188763 & 0.000000 & 107.188763 \\
 & 1.0 & -10.891664 & 108.916638 & 0.000000 & 108.916638 \\
 & 2.5 & -11.150845 & 111.508451 & 0.000000 & 111.508451 \\
 & 5.0 & -11.582814 & 115.828138 & 0.000000 & 115.828138 \\
\hline\hline
\end{tabular}
\end{table*}

A particularly relevant feature emerges when comparing the optical transitions corresponding to $\Delta m = -1$ (Fig.~\ref{fig:optics}) and $\Delta m = +1$ (Fig.~\ref{fig:optical_properties}). A pronounced asymmetry in the transition energies is observed, a phenomenon directly attributable to the AB flux. This difference reflects the symmetry-breaking induced by the coupling between the electron's angular momentum and the vector potential, an intrinsic mechanism in mesoscopic systems under quantum confinement.

Figures ~\ref{fig:optics} and \ref{fig:optical_properties} show that the resonance associated with the ($\Delta m = -1$) transition occurs at a substantially higher photon energy than that of the ($\Delta m = +1$) transition. This imbalance, quantified in Table \ref{tab:comparacao_fluxo_completa} for ($\ell = 0.1$), highlights the high sensitivity of the optical spectrum to the boundary conditions imposed by the magnetic flux and, therefore, to the geometric phase acquired by the charge carriers.

The origin of this asymmetry can be interpreted from the structure of the effective potential ($V_{\mathrm{eff}}(\rho)$) (Eq.~\ref{eq:Veff_pureE}), whose dominant term is proportional to ($(m - \ell)^2$), where ($\ell$) is the magnetic flux expressed in units of the quantum flux ($\Phi_0 = h/e$). Both analyzed transitions originate from the ground state ($(n = 0, m = 0)$), whose energy is modulated by ($\ell^2$). For a non-zero flux ($(\ell > 0)$), the final state ($m = -1$) is governed by ($(m - \ell)^2 = (1 + \ell)^2$), while the state ($m = +1$) depends on ($(1 - \ell)^2$). Since ($(1 + \ell)^2 > (1 - \ell)^2$), the AB flux raises the energy level of the ($m = -1$) state and reduces that of the ($m = +1$) state, originating the observed asymmetry between the transition energies.

The fundamental nature of this effect is confirmed by considering the limiting case ($\ell = 0$), the results of which are also found in Table \ref{tab:comparacao_fluxo_completa}. In the absence of magnetic flux, the ($n = 0, m = 0$) and ($n = 0, m = 1$) states become degenerate, as predicted analytically (Eq.~\ref{eq:energy_eigenvalues}), and the ($\Delta m = +1$) transition is energetically forbidden, since ($\Delta E(0 \to 1) = 0$), regardless of the values of ($\tau$) or ($B$). On the other hand, the ($\Delta m = -1$) transition remains allowed, presenting a finite energy that increases with the increase of torsion and magnetic field.

In summary, the introduction of the AB flux breaks the degeneracy between the ($m = 0$) and ($m = 1$) states, ``activating'' the ($\Delta m = +1$) transition and allowing both ($\Delta m = \pm 1$) transitions to coexist at distinct optical energies. This behavior constitutes an unequivocal signature of the AB phase in the quantum ring's absorption spectrum, reinforcing the relevance of the coupling between topology, confinement, and external fields in the mesoscopic regime.

To complete the analysis of the transition $\Delta m = -1$ (introduced in Fig.~\ref{fig:optics}), the PCS, $\sigma$, is investigated as a function of the incident photon energy, with the results presented in Fig.~\ref{fig:photoionization}. The analysis considers the same transition from the ground state ($n=0, m=0$) to the first excited state ($n=0, m=-1$). The results reveal two significant effects induced by the torsion. First, a \textit{blueshift} of the resonance peak is observed. As the torsion density $\tau$ increases, the peak of the cross-section shifts to higher energies, reflecting the increase in the transition energy. Second, a pronounced \textit{suppression of the peak amplitude} occurs with increasing $\tau$. This behavior is attributed to a reduction in the dipole transition matrix element, as torsion alters the spatial profiles of the electron wave functions, decreasing their overlap and thus the transition probability.

\section{Oscillator strength and its torsion/topology dependence}
\label{sec:osc_strength}

The (dimensionless) oscillator strength quantifies the probability of an electric-dipole transition and, in the length gauge, is
\begin{equation}
f_{fi}
=\frac{2 m^{*}}{\hbar^{2}}\,(E_f-E_i)\,\big|\langle \psi_f |\, \hat{\boldsymbol r}\!\cdot\!\hat{\boldsymbol e}_r\,| \psi_i \rangle\big|^{2},
\label{eq:def_f_fi}
\end{equation}
where $\hat{\boldsymbol e}_r$ is the radial polarization. For the cylindrically symmetric eigenstates
$\psi_{n,m}(\rho,\varphi,z)=R_{n,m}(\rho)\,e^{im\varphi}\,e^{ik_z z}$,
the dipole matrix element reduces to a purely radial integral with the selection rule $\Delta m=\pm 1$:
\begin{align}
M_{fi}
&=\langle \psi_{n',m\pm 1} |\, \hat{\boldsymbol r}\!\cdot\!\hat{\boldsymbol e}_r \,| \psi_{n,m} \rangle
=\pi\!\int_{0}^{\infty}\! R_{n',m\pm1}(\rho)\, \rho^{2}\, R_{n,m}(\rho)\, d\rho,
\label{eq:Mfi_radial}
\end{align}
with $E_f-E_i\equiv\Delta E$ given by Eq.~\eqref{eq:energy_eigenvalues}. Using the normalized eigenfunctions in Eq.~\eqref{eq:wavefunction_explicit} and the change of variable $\xi=|\Lambda|\rho^{2}$, the integral in \eqref{eq:Mfi_radial} can be written in the compact, dimensionless form
\begin{align}
M_{fi}
&=\frac{\pi}{2\,|\Lambda|^{3/2}}\,
\mathcal{N}_{n',\alpha_f}\,\mathcal{N}_{n,\alpha_i}
\!\int_{0}^{\infty}\! \xi^{\frac{\alpha_i+\alpha_f}{2}+\tfrac{1}{2}}
e^{-\xi}\,
L_{n'}^{(\alpha_f)}(\xi)\,L_{n}^{(\alpha_i)}(\xi)\, d\xi,
\label{eq:Mfi_dimless}
\end{align}
where $\alpha_i=|m-l|$, $\alpha_f=|m\!\pm\!1-l|$, and the normalization constant is
$\mathcal{N}_{n,\alpha}=\sqrt{2|\Lambda|\,n!/\Gamma(n+\alpha+1)}$.
The remaining integral is a standard Laguerre overlap (readily evaluated analytically from tabulated Kummer-Laguerre identities or numerically by quadrature). Substituting \eqref{eq:Mfi_dimless} into \eqref{eq:def_f_fi} yields $f_{fi}$ for any allowed $(n,m)\!\to\!(n',m\pm1)$.
\begin{figure}[tbhp]
  \centering
  \includegraphics[width=\linewidth]{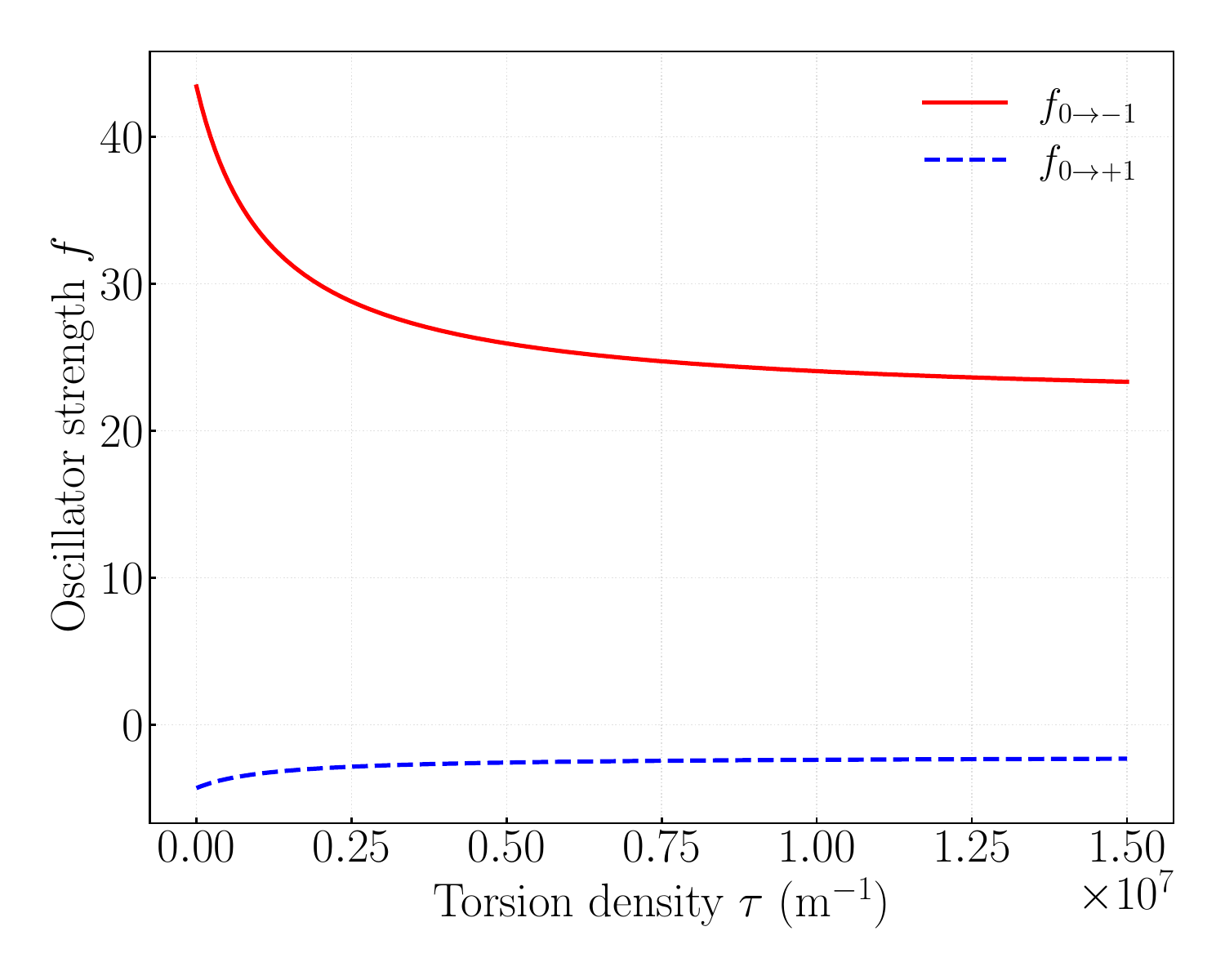}
  \caption{Oscillator strengths for $(n,m)=(0,0)\to(0,\pm1)$ versus the torsion density $\tau$, at $B=5~\mathrm{T}$, $k_z=\pi\times10^{9}~\mathrm{m^{-1}}$, and $l=0.1$. The trend reflects the competition $f_{fi}\propto\Delta E\,|M_{fi}|^{2}$: increasing $\tau$ blueshifts levels (larger $\Delta E$) but compresses wavefunctions (smaller $|M_{fi}|$).}
  \label{fig:fosc_tau}
\end{figure}
\begin{figure}[tbhp]
  \centering
  \includegraphics[width=\linewidth]{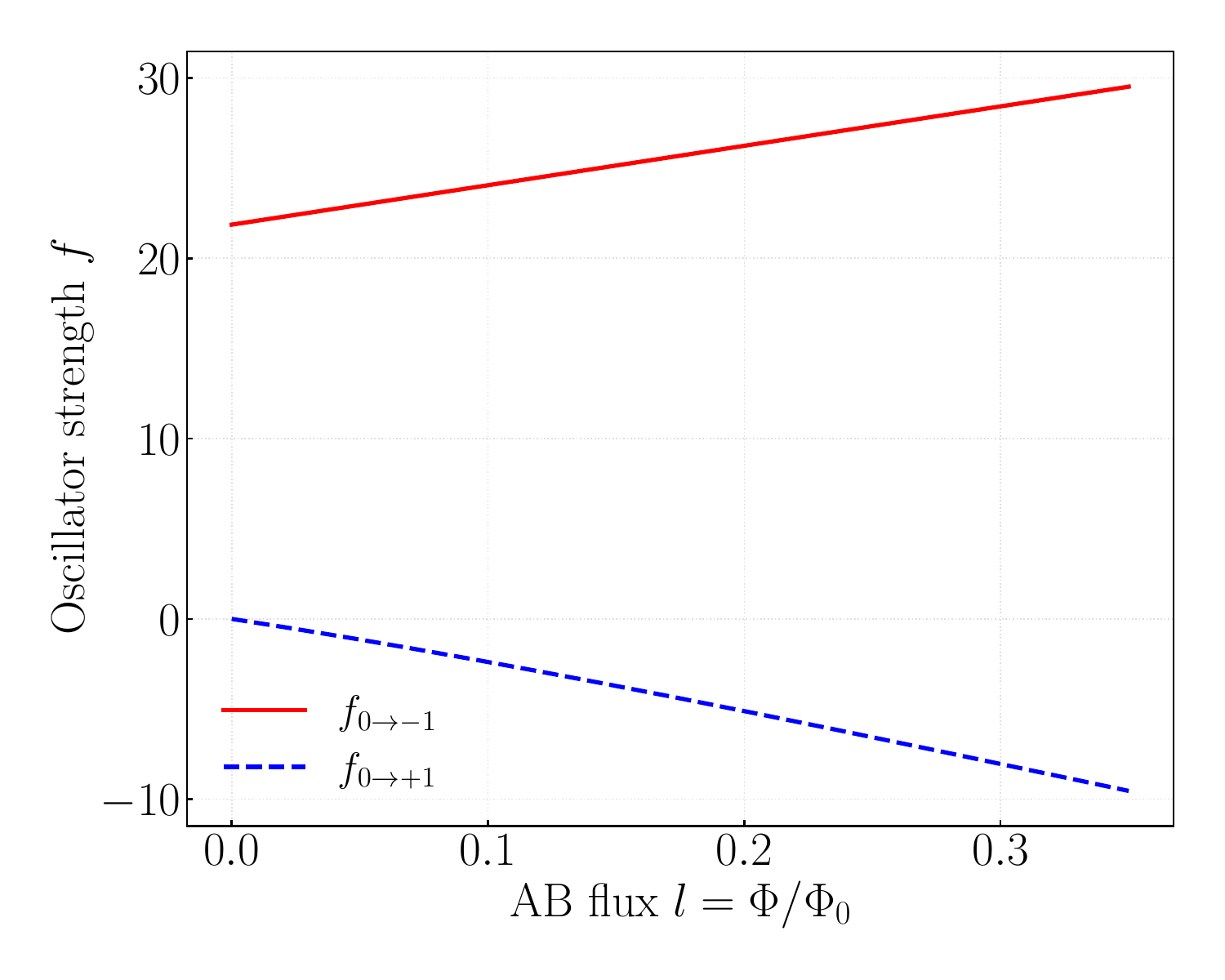}
  \caption{AB-induced splitting of oscillator strengths: $f_{0\to\pm1}$ versus $l=\Phi/\Phi_0$ for $\tau=10^{7}~\mathrm{m^{-1}}$, $B=5~\mathrm{T}$, and $k_z=\pi\times10^{9}~\mathrm{m^{-1}}$. The AB phase breaks the $m\leftrightarrow -m$ symmetry, yielding distinct optical weights for the two angular channels.}
  \label{fig:fosc_split}
\end{figure}
Two mechanisms control $f_{fi}$: (i) the detuning $\Delta E$, which increases with the effective frequency $\omega_{\tau}+\omega_c$ [see Eq.~\eqref{eq:energy_eigenvalues}], and (ii) the radial overlap $|M_{fi}|$, which decreases as torsion and magnetic field increase the confinement scale $|\Lambda|=|k_z\tau+eB/2\hbar|$ and compress the wavefunctions. Since $M_{fi}\propto |\Lambda|^{-1/2}$ times a Laguerre overlap that is further reduced as $|\Lambda|$ grows, torsion ($\tau$) and field ($B$) tend to increase $\Delta E$ but decrease $|M_{fi}|$. The observed $f_{fi}$ thus reflects a competition between these tendencies, whose net outcome depends on $k_z$, $\tau$, $B$, and the AB parameter $l$.

A finite AB flux $l$ breaks the $m\!\leftrightarrow\!-m$ symmetry through the $(m-l)^2$ dependence in both energies and eigenfunctions. For transitions from $m=0$ one generally finds $\Delta E_{0\to-1}>\Delta E_{0\to+1}$ and $|M_{0\to-1}|<|M_{0\to+1}|$ for $l>0$, which produces clearly split oscillator strengths $f_{0\to\pm1}$. This constitutes a direct interferometric signature of the AB phase in the optical weights of the $\Delta m=\pm1$ channels.

Within the linear response used in Sec.~\ref{sec:optical_properties_framework}, the integrated area of an absorption line and the amplitude of the dispersive refractive-index change scale with $f_{fi}$ (for fixed broadening), so the same torsion/flux parameters that tune $\Delta E$ and $|M_{fi}|$ also reweight the optical line strengths. In the nonlinear regime, $M_{fi}$ determines the magnitude of third-order corrections, thereby affecting saturation and the onset of optical switching.

Figure~\ref{fig:fosc_tau} displays $f_{0\to\pm1}$ versus $\tau$ at fixed $B=5~\mathrm{T}$, $k_z=\pi\times10^{9}~\mathrm{m^{-1}}$, and $l=0.1$. The $0\to-1$ channel strengthens with increasing torsion across the explored range, while the $0\to+1$ branch remains weaker, consistent with a torsion-driven growth of $\Delta E$ outweighing the overlap reduction for the $m'=-1$ final state. Figure~\ref{fig:fosc_split} shows $f_{0\to\pm1}$ versus $l$ for $\tau=10^{7}~\mathrm{m^{-1}}$ and $B=5~\mathrm{T}$, revealing the AB-induced splitting $f_{0\to-1}>f_{0\to+1}$ for $l>0$. Together, these panels highlight how geometry (torsion), magnetic quantization, and topology (AB phase) provide complementary parameters to tailor both transition energies and line areas via $f_{fi}$.

\section{Conclusion}
\label{conclusion}

We analyzed the linear and third-order nonlinear optical responses, as well as the photoionization cross section, of a confined charged particle in a non-Euclidean background with uniform torsion. Using the Laplace-Beltrami formalism with minimal coupling to a uniform magnetic field and an Aharonov-Bohm flux, we derived a radial equation in Schrödinger form with a clear effective potential; its exact Laguerre-Kummer solutions provided closed-form wavefunctions and spectra, which we used to compute absorption, refractive-index changes, and photoionization. Torsion acted as an effective confinement parameter, widening level spacings, producing a systematic blueshift of resonances, and reducing radial overlap, thereby weakening dipole matrix elements and diminishing both linear peaks and nonlinear dispersive features. At high intensity, a negative $\alpha^{(3)}$ enabled optical switching (gain near resonance), with threshold and spectral shape strongly modulated by $\tau$ and the AB flux. For photoionization, torsion and AB flux jointly tuned the resonance energy and PCS amplitude, consistent with the modified bound-state profiles and the interferometric phase accumulated around the ring~\cite{aharonov1959significance,PRB.1996.53.6947,PhysRevB.96.165303,WenfangXie2013,pereira2024rotating}. Together with magnetic quantization (Landau/Fock-Darwin physics)~\cite{PLA.2016.380.3847,SPC.2022.5.029,AdP.2022.535.2200371}, these effects established a coherent picture of geometric-topological control over optical processes in mesoscopic systems.

The framework can be extended along several directions: (a) incorporating spin-orbit coupling and Zeeman terms to explore spin-selective selection rules and AB-driven spin interferometry \cite{PRB.2005.71.033309}; (b) treating anisotropic or position-dependent effective mass and strain-induced gauge fields to connect with elastic/curved Landau-level engineering \cite{SPC.2022.5.029,elastic,elasticll}; (c) analyzing disorder and finite-temperature damping to compare with magneto-transport and magneto-caloric signatures under AB flux \cite{JLTP.2021.202.83,PB.2024.673.415438}; (d) extending the optical theory to higher orders and pump-probe geometries and to impurity-bound-to-continuum transitions under strong laser dressing \cite{PhysRevB.105.075113,RevModPhys.83.543}. These avenues underscore the breadth of device concepts, from flux-tunable modulators to geometry, programmed gain elements, available when torsion, AB phase, and magnetic quantization are treated on equal footing.

A relevant experimental question that emerges from our results is how to distinguish the contribution of torsion ($\tau$) from that of the magnetic field ($B$), given that both effectively increase confinement and shift the peaks of the optical properties.

Our theoretical model provides a clear experimental signature to isolate the torsional effect. As derived in Eq.~\ref{eq:energy_eigenvalues}, the energy spectrum depends on the torsion frequency, $\omega_\tau$, and the cyclotron frequency, $\omega_c$. The fundamental distinction is that the torsion effect ($\omega_\tau$) is directly proportional to the electron's momentum $k_z$ along the $z$-axis, vanishing if $k_z = 0$.

This suggests a direct experiment: by keeping the magnetic field $B$ constant, any modulation in the absorption energy (the blueshift) that correlates with a variation in $k_z$ would be an unequivocal signature of the torsion $\tau$.

\appendix

\section{Wave Function Normalization}
\label{app:normalization}

The normalization of the wave function is imposed by the condition that the probability of finding the particle in the entire two-dimensional space is unity. In cylindrical coordinates, the condition is:
\begin{equation}
 \int_0^\infty \abs{R_{n,m}(\rho)}^2 \rho \, d\rho = 1.
 \label{eq:norm_condition_app}
\end{equation}
To solve this integral, we substitute the unnormalized wave function into Eq.~\eqref{eq:norm_condition_app} and perform the change of variables from $\rho$ to $\xi$. Knowing that $\xi = \abs{\Lambda}\rho^2$, we have $\rho = \sqrt{\xi/\abs{\Lambda}}$ and $d\rho = d\xi / (2\sqrt{\abs{\Lambda}\xi})$. Therefore, the integration element becomes $\rho \, d\rho = d\xi / (2\abs{\Lambda})$. The normalization integral takes the form:
\begin{align}
 1 &= \abs{C}^2 \int_0^\infty \left( \xi^{\frac{\abs{m-l}}{2}} e^{-\xi/2} L_n^{\abs{m-l}}(\xi) \right)^2 \frac{d\xi}{2\abs{\Lambda}} \\
 1 &= \frac{\abs{C}^2}{2\abs{\Lambda}} \int_0^\infty \xi^{\abs{m-l}} e^{-\xi} \left( L_n^{\abs{m-l}}(\xi) \right)^2 d\xi.
\end{align}
The remaining integral is a standard form of the orthogonality relation for the generalized Laguerre polynomials:
\begin{equation}
 \int_0^\infty x^p e^{-x} \left[ L_n^p(x) \right]^2 dx = \frac{\Gamma(n+p+1)}{n!},
\end{equation}
where $\Gamma(z)$ is the gamma function. Identifying $p = \abs{m-l}$, we obtain:
\begin{equation}
 1 = \frac{\abs{C}^2}{2\abs{\Lambda}} \frac{\Gamma(n+\abs{m-l}+1)}{n!}.
\end{equation}
Isolating the normalization constant $C$, we find:
\begin{equation}
 C = \sqrt{\frac{2\abs{\Lambda} \, n!}{\Gamma(n+\abs{m-l}+1)}}.
\end{equation}
Finally, the properly normalized radial wave function is:
\begin{equation}
R_{n,m}(\rho) = \sqrt{\frac{2\abs{\Lambda} \, n!}{\Gamma(n+\abs{m-l}+1)}} \, \xi^{\frac{\abs{m-l}}{2}} e^{-\xi/2} L_n^{\abs{m-l}}(\xi).
\label{eq:normalized_wf_app}
\end{equation}
This expression is used to calculate the dipole matrix elements that govern the system's nonlinear optical properties.

\section*{Acknowledgments}

This work was supported by CAPES (Finance Code 001), CNPq (Grant 306308/2022-3), and FAPEMA (Grants UNIVERSAL-06395/22 and APP-12256/22).

\bibliographystyle{model1a-num-names}
%\bibliography{References}

\begin{thebibliography}{118}
\expandafter\ifx\csname natexlab\endcsname\relax\def\natexlab#1{#1}\fi
\providecommand{\bibinfo}[2]{#2}
\ifx\xfnm\relax \def\xfnm[#1]{\unskip,\space#1}\fi
%Type = Article
\bibitem[{Meunier et~al.(2016)Meunier, Souza~Filho, Barros, and
  Dresselhaus}]{RevModPhys.88.025005}
\bibinfo{author}{V.~Meunier}, \bibinfo{author}{A.~G. Souza~Filho},
  \bibinfo{author}{E.~B. Barros}, \bibinfo{author}{M.~S. Dresselhaus},
  \bibinfo{journal}{Rev. Mod. Phys.} \bibinfo{volume}{88}
  (\bibinfo{year}{2016}) \bibinfo{pages}{025005}.
%Type = Article
\bibitem[{Bryant(1989)}]{PhysRevB.40.1620}
\bibinfo{author}{G.~W. Bryant}, \bibinfo{journal}{Phys. Rev. B}
  \bibinfo{volume}{40} (\bibinfo{year}{1989}) \bibinfo{pages}{1620--1629}.
%Type = Article
\bibitem[{Khlebnikov and Hillhouse(2009)}]{PhysRevB.80.115316}
\bibinfo{author}{S.~Khlebnikov}, \bibinfo{author}{H.~W. Hillhouse},
  \bibinfo{journal}{Phys. Rev. B} \bibinfo{volume}{80} (\bibinfo{year}{2009})
  \bibinfo{pages}{115316}.
%Type = Article
\bibitem[{Roodenko et~al.(2010)Roodenko, Goldthorpe, McIntyre, and
  Chabal}]{PhysRevB.82.115210}
\bibinfo{author}{K.~Roodenko}, \bibinfo{author}{I.~A. Goldthorpe},
  \bibinfo{author}{P.~C. McIntyre}, \bibinfo{author}{Y.~J. Chabal},
  \bibinfo{journal}{Phys. Rev. B} \bibinfo{volume}{82} (\bibinfo{year}{2010})
  \bibinfo{pages}{115210}.
%Type = Article
\bibitem[{Sidorova et~al.(2024)Sidorova, Semenov, Zaccone, Charaev, Gonzalez,
  Schilling, Gyger, and Steinhauer}]{PhysRevB.110.134513}
\bibinfo{author}{M.~Sidorova}, \bibinfo{author}{A.~D. Semenov},
  \bibinfo{author}{A.~Zaccone}, \bibinfo{author}{I.~Charaev},
  \bibinfo{author}{M.~Gonzalez}, \bibinfo{author}{A.~Schilling},
  \bibinfo{author}{S.~Gyger}, \bibinfo{author}{S.~Steinhauer},
  \bibinfo{journal}{Phys. Rev. B} \bibinfo{volume}{110} (\bibinfo{year}{2024})
  \bibinfo{pages}{134513}.
%Type = Article
\bibitem[{Koo et~al.(2024)Koo, Moon, Kang, Joo, Lee, Lee, Kravtsov, and
  Park}]{LSA.2024.13.30}
\bibinfo{author}{Y.~Koo}, \bibinfo{author}{T.~Moon}, \bibinfo{author}{M.~Kang},
  \bibinfo{author}{H.~Joo}, \bibinfo{author}{C.~Lee}, \bibinfo{author}{H.~Lee},
  \bibinfo{author}{V.~Kravtsov}, \bibinfo{author}{K.-D. Park},
  \bibinfo{journal}{Light: Science {\&} Applications} \bibinfo{volume}{13}
  (\bibinfo{year}{2024}) \bibinfo{pages}{30}.
%Type = Article
\bibitem[{Ban et~al.(2023)Ban, Nie, Lei, Yi, Vinu, Bao, and
  Liu}]{MRL.2023.11.21}
\bibinfo{author}{S.~Ban}, \bibinfo{author}{X.~Nie}, \bibinfo{author}{Z.~Lei},
  \bibinfo{author}{J.~Yi}, \bibinfo{author}{A.~Vinu}, \bibinfo{author}{Y.~Bao},
  \bibinfo{author}{Y.~Liu}, \bibinfo{journal}{Materials Research Letters}
  \bibinfo{volume}{11} (\bibinfo{year}{2023}) \bibinfo{pages}{21--52}.
%Type = Misc
\bibitem[{Levine(1993)}]{Levine1993QWIP}
\bibinfo{author}{B.~F. Levine}, \bibinfo{title}{Quantum-well infrared
  photodetectors}, \bibinfo{year}{1993}.
%Type = Article
\bibitem[{Miller et~al.(1984)Miller, Chemla, Damen, Gossard, Wiegmann, Wood,
  and Burrus}]{Miller1984QCSE_PRL}
\bibinfo{author}{D.~A.~B. Miller}, \bibinfo{author}{D.~S. Chemla},
  \bibinfo{author}{T.~C. Damen}, \bibinfo{author}{A.~C. Gossard},
  \bibinfo{author}{W.~Wiegmann}, \bibinfo{author}{T.~H. Wood},
  \bibinfo{author}{C.~A. Burrus}, \bibinfo{journal}{Physical Review Letters}
  \bibinfo{volume}{53} (\bibinfo{year}{1984}) \bibinfo{pages}{2173--2176}.
%Type = Article
\bibitem[{Miller et~al.(1986)Miller, Chemla, Damen, Gossard, Wiegmann, and
  Burrus}]{Miller1986QCSE_Devices}
\bibinfo{author}{D.~A.~B. Miller}, \bibinfo{author}{D.~S. Chemla},
  \bibinfo{author}{T.~C. Damen}, \bibinfo{author}{A.~C. Gossard},
  \bibinfo{author}{W.~Wiegmann}, \bibinfo{author}{C.~A. Burrus},
  \bibinfo{journal}{IEEE Journal of Quantum Electronics} \bibinfo{volume}{22}
  (\bibinfo{year}{1986}) \bibinfo{pages}{1816--1830}.
%Type = Article
\bibitem[{Konstantatos and Sargent(2010)}]{Konstantatos2010CQD_PD_Review}
\bibinfo{author}{G.~Konstantatos}, \bibinfo{author}{E.~H. Sargent},
  \bibinfo{journal}{Nature Nanotechnology} \bibinfo{volume}{5}
  (\bibinfo{year}{2010}) \bibinfo{pages}{391--400}.
%Type = Article
\bibitem[{Rogalski et~al.(2002)Rogalski, Kopytko, and
  Martyniuk}]{Rogalski2002QDIP_Review}
\bibinfo{author}{A.~Rogalski}, \bibinfo{author}{M.~Kopytko},
  \bibinfo{author}{P.~Martyniuk}, \bibinfo{journal}{Progress in Quantum
  Electronics} \bibinfo{volume}{27} (\bibinfo{year}{2002})
  \bibinfo{pages}{59--210}.
%Type = Article
\bibitem[{Yan et~al.(2009)Yan, Gargas, and Yang}]{Yan2009NanowirePhotonics}
\bibinfo{author}{R.~Yan}, \bibinfo{author}{D.~Gargas},
  \bibinfo{author}{P.~Yang}, \bibinfo{journal}{Nature Photonics}
  \bibinfo{volume}{3} (\bibinfo{year}{2009}) \bibinfo{pages}{569--576}.
%Type = Article
\bibitem[{Arakawa and Sakaki(1982)}]{Arakawa1982QDlaser_APL}
\bibinfo{author}{Y.~Arakawa}, \bibinfo{author}{H.~Sakaki},
  \bibinfo{journal}{Applied Physics Letters} \bibinfo{volume}{40}
  (\bibinfo{year}{1982}) \bibinfo{pages}{939--941}.
%Type = Article
\bibitem[{Wang et~al.(2019)Wang, Liu, Johnson, Wong-Leung, Jagadish, Livshits,
  Lee, and Bowers}]{Wang2019QD_SiPhot}
\bibinfo{author}{Z.~Wang}, \bibinfo{author}{A.~Y. Liu}, \bibinfo{author}{S.~G.
  Johnson}, \bibinfo{author}{J.~Wong-Leung}, \bibinfo{author}{C.~Jagadish},
  \bibinfo{author}{D.~Livshits}, \bibinfo{author}{C.~Lee},
  \bibinfo{author}{J.~E. Bowers}, \bibinfo{journal}{Nature Photonics}
  \bibinfo{volume}{13} (\bibinfo{year}{2019}) \bibinfo{pages}{45--49}.
%Type = Article
\bibitem[{Lorke et~al.(2000)Lorke, Luyken, Govorov, Kotthaus, Garcia, and
  Petroff}]{Lorke2000QR_PRL}
\bibinfo{author}{A.~Lorke}, \bibinfo{author}{R.~J. Luyken},
  \bibinfo{author}{A.~O. Govorov}, \bibinfo{author}{J.~P. Kotthaus},
  \bibinfo{author}{J.~M. Garcia}, \bibinfo{author}{P.~M. Petroff},
  \bibinfo{journal}{Physical Review Letters} \bibinfo{volume}{84}
  (\bibinfo{year}{2000}) \bibinfo{pages}{2223--2226}.
%Type = Article
\bibitem[{Warburton et~al.(2000)Warburton, Sch{\"a}flein, Haft, Bickel, Lorke,
  Karrai, Garcia, Schoenfeld, and Petroff}]{Warburton2000QD_Nature}
\bibinfo{author}{R.~J. Warburton}, \bibinfo{author}{C.~Sch{\"a}flein},
  \bibinfo{author}{D.~Haft}, \bibinfo{author}{F.~Bickel},
  \bibinfo{author}{A.~Lorke}, \bibinfo{author}{K.~Karrai},
  \bibinfo{author}{J.~M. Garcia}, \bibinfo{author}{W.~Schoenfeld},
  \bibinfo{author}{P.~M. Petroff}, \bibinfo{journal}{Nature}
  \bibinfo{volume}{405} (\bibinfo{year}{2000}) \bibinfo{pages}{926--929}.
%Type = Article
\bibitem[{Yu et~al.(2025)Yu, Chen, Fu, Zhu, and Feng}]{YU2025130856}
\bibinfo{author}{L.~Yu}, \bibinfo{author}{Y.~Chen}, \bibinfo{author}{Z.~Fu},
  \bibinfo{author}{T.~Zhu}, \bibinfo{author}{X.~Feng},
  \bibinfo{journal}{Physics Letters A} \bibinfo{volume}{557}
  (\bibinfo{year}{2025}) \bibinfo{pages}{130856}.
%Type = Article
\bibitem[{Blundo et~al.(2021)Blundo, Cappelluti, Felici, Pettinari, and
  Polimeni}]{10.1063/5.0037852}
\bibinfo{author}{E.~Blundo}, \bibinfo{author}{E.~Cappelluti},
  \bibinfo{author}{M.~Felici}, \bibinfo{author}{G.~Pettinari},
  \bibinfo{author}{A.~Polimeni}, \bibinfo{journal}{Applied Physics Reviews}
  \bibinfo{volume}{8} (\bibinfo{year}{2021}) \bibinfo{pages}{021318}.
%Type = Article
\bibitem[{Song et~al.(2024)Song, Ye, Su, Wei, Chen, Liu, Zheng, and
  Hao}]{PhysRevB.109.094105}
\bibinfo{author}{L.~Song}, \bibinfo{author}{R.~Ye}, \bibinfo{author}{C.~Su},
  \bibinfo{author}{C.~Wei}, \bibinfo{author}{D.~Chen},
  \bibinfo{author}{X.~Liu}, \bibinfo{author}{X.~Zheng},
  \bibinfo{author}{H.~Hao}, \bibinfo{journal}{Phys. Rev. B}
  \bibinfo{volume}{109} (\bibinfo{year}{2024}) \bibinfo{pages}{094105}.
%Type = Article
\bibitem[{Soriano et~al.(2024)Soriano, Marian, Dubey, and
  Fiori}]{PhysRevB.109.115434}
\bibinfo{author}{D.~Soriano}, \bibinfo{author}{D.~Marian},
  \bibinfo{author}{P.~Dubey}, \bibinfo{author}{G.~Fiori},
  \bibinfo{journal}{Phys. Rev. B} \bibinfo{volume}{109} (\bibinfo{year}{2024})
  \bibinfo{pages}{115434}.
%Type = Article
\bibitem[{Li et~al.(2024)Li, Wei, Liu, Tang, and Jiang}]{Li2024}
\bibinfo{author}{S.~Li}, \bibinfo{author}{K.~Wei}, \bibinfo{author}{Q.~Liu},
  \bibinfo{author}{Y.~Tang}, \bibinfo{author}{T.~Jiang},
  \bibinfo{journal}{Frontiers of Physics} \bibinfo{volume}{19}
  (\bibinfo{year}{2024}) \bibinfo{pages}{42501}.
%Type = Article
\bibitem[{Katanaev and Volovich(1992)}]{Katanaev1992AnnPhys}
\bibinfo{author}{M.~O. Katanaev}, \bibinfo{author}{I.~V. Volovich},
  \bibinfo{journal}{Annals of Physics} \bibinfo{volume}{216}
  (\bibinfo{year}{1992}) \bibinfo{pages}{1--28}.
%Type = Book
\bibitem[{Kleinert(1989)}]{Kleinert1989GaugeFields}
\bibinfo{author}{H.~Kleinert}, \bibinfo{title}{Gauge Fields in Condensed
  Matter}, volume~\bibinfo{volume}{1}, \bibinfo{publisher}{World Scientific},
  \bibinfo{address}{Singapore}, \bibinfo{year}{1989}.
%Type = Article
\bibitem[{Nelson(1987)}]{Nelson1987DefectsGeometry}
\bibinfo{author}{D.~R. Nelson}, \bibinfo{journal}{Physics Today}
  \bibinfo{volume}{40} (\bibinfo{year}{1987}) \bibinfo{pages}{32--38}.
%Type = Article
\bibitem[{Jensen and Koppe(1971)}]{Jensen1991}
\bibinfo{author}{H.~Jensen}, \bibinfo{author}{H.~Koppe},
  \bibinfo{journal}{Annals of Physics} \bibinfo{volume}{63}
  (\bibinfo{year}{1971}) \bibinfo{pages}{586--591}.
%Type = Article
\bibitem[{Puntigam and Soleng(1997)}]{Puntigam1997CQG}
\bibinfo{author}{R.~A. Puntigam}, \bibinfo{author}{H.~H. Soleng},
  \bibinfo{journal}{Classical and Quantum Gravity} \bibinfo{volume}{14}
  (\bibinfo{year}{1997}) \bibinfo{pages}{1129--1149}.
%Type = Article
\bibitem[{Katanaev(2005)}]{Katanaev2005PhysUsp}
\bibinfo{author}{M.~O. Katanaev}, \bibinfo{journal}{Physics-Uspekhi}
  \bibinfo{volume}{48} (\bibinfo{year}{2005}) \bibinfo{pages}{675--701}.
%Type = Article
\bibitem[{de~A.~Marques et~al.(2001)de~A.~Marques, Furtado, Bezerra, and
  Moraes}]{Furtado1999PLA}
\bibinfo{author}{G.~de~A.~Marques}, \bibinfo{author}{C.~Furtado},
  \bibinfo{author}{V.~B. Bezerra}, \bibinfo{author}{F.~Moraes},
  \bibinfo{journal}{Journal of Physics A: Mathematical and General}
  \bibinfo{volume}{34} (\bibinfo{year}{2001}) \bibinfo{pages}{5945--5954}.
%Type = Article
\bibitem[{Moraes(2000)}]{Moraes2000IJMPA}
\bibinfo{author}{F.~Moraes}, \bibinfo{journal}{International Journal of Modern
  Physics A} \bibinfo{volume}{15} (\bibinfo{year}{2000})
  \bibinfo{pages}{3943--3952}.
%Type = Article
\bibitem[{B{\"u}ttiker et~al.(1983)B{\"u}ttiker, Imry, and
  Landauer}]{Buttiker1983PLA}
\bibinfo{author}{M.~B{\"u}ttiker}, \bibinfo{author}{Y.~Imry},
  \bibinfo{author}{R.~Landauer}, \bibinfo{journal}{Physics Letters A}
  \bibinfo{volume}{96} (\bibinfo{year}{1983}) \bibinfo{pages}{365--367}.
%Type = Article
\bibitem[{Gefen et~al.(1984)Gefen, Imry, and Azbel}]{Gefen1984PRL}
\bibinfo{author}{Y.~Gefen}, \bibinfo{author}{Y.~Imry}, \bibinfo{author}{M.~Y.
  Azbel}, \bibinfo{journal}{Physical Review Letters} \bibinfo{volume}{52}
  (\bibinfo{year}{1984}) \bibinfo{pages}{129--132}.
%Type = Article
\bibitem[{Webb et~al.(1985)Webb, Washburn, Umbach, and Laibowitz}]{Webb1985PRL}
\bibinfo{author}{R.~A. Webb}, \bibinfo{author}{S.~Washburn},
  \bibinfo{author}{C.~P. Umbach}, \bibinfo{author}{R.~B. Laibowitz},
  \bibinfo{journal}{Physical Review Letters} \bibinfo{volume}{54}
  (\bibinfo{year}{1985}) \bibinfo{pages}{2696--2699}.
%Type = Article
\bibitem[{Mailly et~al.(1993)Mailly, Chapelier, and Benoit}]{Mailly1993PRL}
\bibinfo{author}{D.~Mailly}, \bibinfo{author}{C.~Chapelier},
  \bibinfo{author}{A.~Benoit}, \bibinfo{journal}{Physical Review Letters}
  \bibinfo{volume}{70} (\bibinfo{year}{1993}) \bibinfo{pages}{2020--2023}.
%Type = Article
\bibitem[{Lorke et~al.(2000)Lorke, Luyken, Govorov, Kotthaus, Garcia, and
  Petroff}]{Lorke2000PRL}
\bibinfo{author}{A.~Lorke}, \bibinfo{author}{R.~J. Luyken},
  \bibinfo{author}{A.~O. Govorov}, \bibinfo{author}{J.~P. Kotthaus},
  \bibinfo{author}{J.~M. Garcia}, \bibinfo{author}{P.~M. Petroff},
  \bibinfo{journal}{Physical Review Letters} \bibinfo{volume}{84}
  (\bibinfo{year}{2000}) \bibinfo{pages}{2223--2226}.
%Type = Article
\bibitem[{Reimann and Manninen(2002)}]{Reimann2002RMP}
\bibinfo{author}{S.~M. Reimann}, \bibinfo{author}{M.~Manninen},
  \bibinfo{journal}{Reviews of Modern Physics} \bibinfo{volume}{74}
  (\bibinfo{year}{2002}) \bibinfo{pages}{1283--1342}.
%Type = Article
\bibitem[{Viefers et~al.(2004)Viefers, Koskinen, Saarikoski, and
  Manninen}]{Viefers2004PhysE}
\bibinfo{author}{S.~Viefers}, \bibinfo{author}{P.~Koskinen},
  \bibinfo{author}{P.~Saarikoski}, \bibinfo{author}{M.~Manninen},
  \bibinfo{journal}{Physica E: Low-dimensional Systems and Nanostructures}
  \bibinfo{volume}{21} (\bibinfo{year}{2004}) \bibinfo{pages}{1--35}.
%Type = Article
\bibitem[{Sheng et~al.(2002)Sheng, Liu, and Hawrylak}]{Sheng2002PRB}
\bibinfo{author}{W.~Sheng}, \bibinfo{author}{R.~J. Liu},
  \bibinfo{author}{P.~Hawrylak}, \bibinfo{journal}{Physical Review B}
  \bibinfo{volume}{66} (\bibinfo{year}{2002}) \bibinfo{pages}{165316}.
%Type = Article
\bibitem[{Fomin et~al.(2003)Fomin, Gladilin, Devreese, Klimin, Misko, Gijs, and
  Borghs}]{Fomin2003PRB}
\bibinfo{author}{V.~M. Fomin}, \bibinfo{author}{V.~N. Gladilin},
  \bibinfo{author}{J.~T. Devreese}, \bibinfo{author}{S.~N. Klimin},
  \bibinfo{author}{V.~R. Misko}, \bibinfo{author}{M.~A.~M. Gijs},
  \bibinfo{author}{G.~Borghs}, \bibinfo{journal}{Physical Review B}
  \bibinfo{volume}{68} (\bibinfo{year}{2003}) \bibinfo{pages}{075307}.
%Type = Book
\bibitem[{Fomin(2014)}]{Fomin2014Book}
\bibinfo{editor}{V.~M. Fomin} (Ed.), \bibinfo{title}{Physics of Quantum Rings},
  \bibinfo{publisher}{Springer}, \bibinfo{address}{Berlin, Heidelberg},
  \bibinfo{year}{2014}.
%Type = Article
\bibitem[{Aharonov and Bohm(1959)}]{Aharonov1959PR}
\bibinfo{author}{Y.~Aharonov}, \bibinfo{author}{D.~Bohm},
  \bibinfo{journal}{Physical Review} \bibinfo{volume}{115}
  (\bibinfo{year}{1959}) \bibinfo{pages}{485--491}.
%Type = Article
\bibitem[{Aharonov and Bohm(1961)}]{Aharonov1961PR}
\bibinfo{author}{Y.~Aharonov}, \bibinfo{author}{D.~Bohm},
  \bibinfo{journal}{Physical Review} \bibinfo{volume}{123}
  (\bibinfo{year}{1961}) \bibinfo{pages}{1511--1524}.
%Type = Article
\bibitem[{Chambers(1960)}]{Chambers1960PRL}
\bibinfo{author}{R.~G. Chambers}, \bibinfo{journal}{Physical Review Letters}
  \bibinfo{volume}{5} (\bibinfo{year}{1960}) \bibinfo{pages}{3--5}.
%Type = Article
\bibitem[{Tonomura et~al.(1986)Tonomura, Osakabe, Matsuda, Kawasaki, Endo,
  Yano, and Yamada}]{Tonomura1986Nature}
\bibinfo{author}{A.~Tonomura}, \bibinfo{author}{N.~Osakabe},
  \bibinfo{author}{T.~Matsuda}, \bibinfo{author}{T.~Kawasaki},
  \bibinfo{author}{J.~Endo}, \bibinfo{author}{S.~Yano},
  \bibinfo{author}{H.~Yamada}, \bibinfo{journal}{Nature} \bibinfo{volume}{324}
  (\bibinfo{year}{1986}) \bibinfo{pages}{524--527}.
%Type = Article
\bibitem[{Washburn and Webb(1992)}]{Washburn1992RPP}
\bibinfo{author}{S.~Washburn}, \bibinfo{author}{R.~A. Webb},
  \bibinfo{journal}{Reports on Progress in Physics} \bibinfo{volume}{55}
  (\bibinfo{year}{1992}) \bibinfo{pages}{1311--1383}.
%Type = Article
\bibitem[{L{\'e}vy et~al.(1990)L{\'e}vy, Dolan, Dunsmuir, and
  Bouchiat}]{Levy1990PRL}
\bibinfo{author}{L.~P. L{\'e}vy}, \bibinfo{author}{G.~Dolan},
  \bibinfo{author}{J.~Dunsmuir}, \bibinfo{author}{H.~Bouchiat},
  \bibinfo{journal}{Physical Review Letters} \bibinfo{volume}{64}
  (\bibinfo{year}{1990}) \bibinfo{pages}{2074--2077}.
%Type = Article
\bibitem[{Chandrasekhar et~al.(1991)Chandrasekhar, Webb, Brady, Ketchen,
  Gallagher, and Kleinsasser}]{Chandrasekhar1991PRL}
\bibinfo{author}{V.~Chandrasekhar}, \bibinfo{author}{R.~A. Webb},
  \bibinfo{author}{M.~J. Brady}, \bibinfo{author}{M.~B. Ketchen},
  \bibinfo{author}{W.~J. Gallagher}, \bibinfo{author}{A.~Kleinsasser},
  \bibinfo{journal}{Physical Review Letters} \bibinfo{volume}{67}
  (\bibinfo{year}{1991}) \bibinfo{pages}{3578--3581}.
%Type = Article
\bibitem[{Cheung et~al.(1988)Cheung, Gefen, Riedel, and Shih}]{Cheung1988PRB}
\bibinfo{author}{H.-F. Cheung}, \bibinfo{author}{Y.~Gefen},
  \bibinfo{author}{E.~K. Riedel}, \bibinfo{author}{W.-H. Shih},
  \bibinfo{journal}{Physical Review B} \bibinfo{volume}{37}
  (\bibinfo{year}{1988}) \bibinfo{pages}{6050--6062}.
%Type = Article
\bibitem[{Landau(1930)}]{Landau1930}
\bibinfo{author}{L.~D. Landau}, \bibinfo{journal}{Zeitschrift f{\"u}r Physik}
  \bibinfo{volume}{64} (\bibinfo{year}{1930}) \bibinfo{pages}{629--637}.
%Type = Article
\bibitem[{Fock(1928)}]{Fock1928}
\bibinfo{author}{V.~Fock}, \bibinfo{journal}{Zeitschrift f{\"u}r Physik}
  \bibinfo{volume}{47} (\bibinfo{year}{1928}) \bibinfo{pages}{446--448}.
%Type = Article
\bibitem[{Darwin(1930)}]{Darwin1930}
\bibinfo{author}{C.~G. Darwin}, \bibinfo{journal}{Proceedings of the Cambridge
  Philosophical Society} \bibinfo{volume}{27} (\bibinfo{year}{1930})
  \bibinfo{pages}{86--90}.
%Type = Article
\bibitem[{Maksym and Chakraborty(1990)}]{Maksym1990PRL}
\bibinfo{author}{P.~A. Maksym}, \bibinfo{author}{T.~Chakraborty},
  \bibinfo{journal}{Physical Review Letters} \bibinfo{volume}{65}
  (\bibinfo{year}{1990}) \bibinfo{pages}{108--111}.
%Type = Article
\bibitem[{McEuen et~al.(1991)McEuen, Foxman, Kinaret, Meirav, Kastner,
  Wingreen, and Wind}]{McEuen1991PRL}
\bibinfo{author}{P.~L. McEuen}, \bibinfo{author}{E.~B. Foxman},
  \bibinfo{author}{J.~Kinaret}, \bibinfo{author}{U.~Meirav},
  \bibinfo{author}{M.~A. Kastner}, \bibinfo{author}{N.~S. Wingreen},
  \bibinfo{author}{S.~J. Wind}, \bibinfo{journal}{Physical Review Letters}
  \bibinfo{volume}{66} (\bibinfo{year}{1991}) \bibinfo{pages}{1926--1929}.
%Type = Article
\bibitem[{Tarucha et~al.(1996)Tarucha, Austing, Honda, van~der Hage, and
  Kouwenhoven}]{Tarucha1996PRL}
\bibinfo{author}{S.~Tarucha}, \bibinfo{author}{D.~G. Austing},
  \bibinfo{author}{T.~Honda}, \bibinfo{author}{R.~J. van~der Hage},
  \bibinfo{author}{L.~P. Kouwenhoven}, \bibinfo{journal}{Physical Review
  Letters} \bibinfo{volume}{77} (\bibinfo{year}{1996})
  \bibinfo{pages}{3613--3616}.
%Type = Article
\bibitem[{Beenakker and van Houten(1991)}]{Beenakker1991SSP}
\bibinfo{author}{C.~W.~J. Beenakker}, \bibinfo{author}{H.~van Houten},
  \bibinfo{journal}{Solid State Physics} \bibinfo{volume}{44}
  (\bibinfo{year}{1991}) \bibinfo{pages}{1--228}.
%Type = Article
\bibitem[{Sheik{-}Bahae et~al.(1990{\natexlab{a}})Sheik{-}Bahae, Said, Wei,
  Hagan, and Stryland}]{SheikBahae1990OL}
\bibinfo{author}{M.~Sheik{-}Bahae}, \bibinfo{author}{A.~A. Said},
  \bibinfo{author}{T.~H. Wei}, \bibinfo{author}{D.~J. Hagan},
  \bibinfo{author}{E.~W.~V. Stryland}, \bibinfo{journal}{Optics Letters}
  \bibinfo{volume}{15} (\bibinfo{year}{1990}{\natexlab{a}})
  \bibinfo{pages}{955--957}.
%Type = Article
\bibitem[{Sheik{-}Bahae et~al.(1990{\natexlab{b}})Sheik{-}Bahae, Said, and
  Stryland}]{SheikBahae1991JOSAB}
\bibinfo{author}{M.~Sheik{-}Bahae}, \bibinfo{author}{A.~A. Said},
  \bibinfo{author}{E.~W.~V. Stryland}, \bibinfo{journal}{Journal of the Optical
  Society of America B} \bibinfo{volume}{9}
  (\bibinfo{year}{1990}{\natexlab{b}}) \bibinfo{pages}{405--410}.
%Type = Article
\bibitem[{Wherrett(1984)}]{Wherrett1984JOSAB}
\bibinfo{author}{B.~S. Wherrett}, \bibinfo{journal}{Journal of the Optical
  Society of America B} \bibinfo{volume}{1} (\bibinfo{year}{1984})
  \bibinfo{pages}{67--72}.
%Type = Article
\bibitem[{Chemla and Shah(2001)}]{ChemlaShah2001Nature}
\bibinfo{author}{D.~S. Chemla}, \bibinfo{author}{J.~Shah},
  \bibinfo{journal}{Nature} \bibinfo{volume}{411} (\bibinfo{year}{2001})
  \bibinfo{pages}{549--557}.
%Type = Article
\bibitem[{Dinu et~al.(2003)Dinu, Quochi, and Garcia}]{Dinu2003APL}
\bibinfo{author}{M.~Dinu}, \bibinfo{author}{F.~Quochi},
  \bibinfo{author}{H.~Garcia}, \bibinfo{journal}{Applied Physics Letters}
  \bibinfo{volume}{82} (\bibinfo{year}{2003}) \bibinfo{pages}{2954--2956}.
%Type = Article
\bibitem[{Autere et~al.(2018)Autere, Jussila, Dai, Wang, Lipsanen, and
  Sun}]{Autere2018AdvMater}
\bibinfo{author}{A.~Autere}, \bibinfo{author}{H.~Jussila},
  \bibinfo{author}{Y.~Dai}, \bibinfo{author}{Y.~Wang},
  \bibinfo{author}{H.~Lipsanen}, \bibinfo{author}{Z.~Sun},
  \bibinfo{journal}{Advanced Materials} \bibinfo{volume}{30}
  (\bibinfo{year}{2018}) \bibinfo{pages}{1705963}.
%Type = Article
\bibitem[{Eggleton et~al.(2011)Eggleton, Luther-Davies, and
  Richardson}]{Eggleton2011NatPhoton}
\bibinfo{author}{B.~J. Eggleton}, \bibinfo{author}{B.~Luther-Davies},
  \bibinfo{author}{K.~Richardson}, \bibinfo{journal}{Nature Photonics}
  \bibinfo{volume}{5} (\bibinfo{year}{2011}) \bibinfo{pages}{141--148}.
%Type = Article
\bibitem[{Kippenberg et~al.(2018)Kippenberg, Gaeta, Lipson, and
  Gorodetsky}]{Kippenberg2018Science}
\bibinfo{author}{T.~J. Kippenberg}, \bibinfo{author}{A.~L. Gaeta},
  \bibinfo{author}{M.~Lipson}, \bibinfo{author}{M.~L. Gorodetsky},
  \bibinfo{journal}{Science} \bibinfo{volume}{361} (\bibinfo{year}{2018})
  \bibinfo{pages}{eaan8083}.
%Type = Article
\bibitem[{Hu et~al.(2001)Hu, Zhu, Li, and Xiong}]{Hu2001PRB}
\bibinfo{author}{H.~Hu}, \bibinfo{author}{J.-L. Zhu}, \bibinfo{author}{D.-J.
  Li}, \bibinfo{author}{J.-J. Xiong}, \bibinfo{journal}{Physical Review B}
  \bibinfo{volume}{63} (\bibinfo{year}{2001}) \bibinfo{pages}{195307}.
%Type = Article
\bibitem[{Govorov et~al.(2002)Govorov, Ulloa, Karrai, and
  Warburton}]{Govorov2002PRB}
\bibinfo{author}{A.~O. Govorov}, \bibinfo{author}{S.~E. Ulloa},
  \bibinfo{author}{K.~Karrai}, \bibinfo{author}{R.~J. Warburton},
  \bibinfo{journal}{Physical Review B} \bibinfo{volume}{66}
  (\bibinfo{year}{2002}) \bibinfo{pages}{081309(R)}.
%Type = Article
\bibitem[{Miller(2010)}]{Miller2010NP}
\bibinfo{author}{D.~A.~B. Miller}, \bibinfo{journal}{Nature Photonics}
  \bibinfo{volume}{4} (\bibinfo{year}{2010}) \bibinfo{pages}{3--5}.
%Type = Article
\bibitem[{Bastard et~al.(1990)Bastard, Conwell, Figueiredo, and
  Ferreira}]{Bastard1990PRB_PCS_QW}
\bibinfo{author}{G.~Bastard}, \bibinfo{author}{E.~M. Conwell},
  \bibinfo{author}{J.~M.~V. Figueiredo}, \bibinfo{author}{R.~Ferreira},
  \bibinfo{journal}{Physical Review B} \bibinfo{volume}{41}
  (\bibinfo{year}{1990}) \bibinfo{pages}{7899--7906}.
%Type = Article
\bibitem[{Bryant(1988)}]{Bryant1988PRB_QD_Impurity}
\bibinfo{author}{G.~W. Bryant}, \bibinfo{journal}{Physical Review B}
  \bibinfo{volume}{37} (\bibinfo{year}{1988}) \bibinfo{pages}{8763--8774}.
%Type = Article
\bibitem[{Sari et~al.(1996)Sari, Rezaei, and Tanatar}]{Sari1996PRB_QD_PCS}
\bibinfo{author}{H.~Sari}, \bibinfo{author}{G.~Rezaei},
  \bibinfo{author}{B.~Tanatar}, \bibinfo{journal}{Physical Review B}
  \bibinfo{volume}{54} (\bibinfo{year}{1996}) \bibinfo{pages}{R5227--R5230}.
%Type = Article
\bibitem[{Ishikawa et~al.(1995)Ishikawa, Shah, Damen, and
  Chemla}]{Ishikawa1995PRB_QW_PCS}
\bibinfo{author}{M.~Ishikawa}, \bibinfo{author}{J.~Shah},
  \bibinfo{author}{T.~C. Damen}, \bibinfo{author}{D.~S. Chemla},
  \bibinfo{journal}{Physical Review B} \bibinfo{volume}{51}
  (\bibinfo{year}{1995}) \bibinfo{pages}{4379--4382}.
%Type = Article
\bibitem[{Szeftel et~al.(1992)Szeftel, Bastard, and
  Voos}]{Szeftel1992PRB_QD_Acceptor_PCS}
\bibinfo{author}{J.~Szeftel}, \bibinfo{author}{G.~Bastard},
  \bibinfo{author}{M.~Voos}, \bibinfo{journal}{Physical Review B}
  \bibinfo{volume}{46} (\bibinfo{year}{1992}) \bibinfo{pages}{3892--3899}.
%Type = Article
\bibitem[{Chakraborty and
  Pietil{\"a}inen(1994)}]{Chakraborty1994PRB_Rings_AB_PCS}
\bibinfo{author}{T.~Chakraborty}, \bibinfo{author}{P.~Pietil{\"a}inen},
  \bibinfo{journal}{Physical Review B} \bibinfo{volume}{50}
  (\bibinfo{year}{1994}) \bibinfo{pages}{8460--8463}.
%Type = Article
\bibitem[{Govorov and Ulloa(2004)}]{Govorov2004PRB_Rings_Exciton_AB}
\bibinfo{author}{A.~O. Govorov}, \bibinfo{author}{S.~E. Ulloa},
  \bibinfo{journal}{Physical Review B} \bibinfo{volume}{69}
  (\bibinfo{year}{2004}) \bibinfo{pages}{125105}.
%Type = Article
\bibitem[{Maksym and Chakraborty(1990)}]{Maksym1990PRB_FD_Optics}
\bibinfo{author}{P.~A. Maksym}, \bibinfo{author}{T.~Chakraborty},
  \bibinfo{journal}{Physical Review Letters} \bibinfo{volume}{65}
  (\bibinfo{year}{1990}) \bibinfo{pages}{108--111}.
%Type = Article
\bibitem[{Lugo et~al.(2006)Lugo, Gaspar-Armenta, and
  Duque}]{Lugo2006PRB_QD_PCS_Field}
\bibinfo{author}{C.~Lugo}, \bibinfo{author}{J.~A. Gaspar-Armenta},
  \bibinfo{author}{C.~A. Duque}, \bibinfo{journal}{Physical Review B}
  \bibinfo{volume}{74} (\bibinfo{year}{2006}) \bibinfo{pages}{205338}.
%Type = Article
\bibitem[{Xie(2013)}]{Xie2013SSC_QR_PCS}
\bibinfo{author}{W.-F. Xie}, \bibinfo{journal}{Solid State Communications}
  \bibinfo{volume}{168} (\bibinfo{year}{2013}) \bibinfo{pages}{6--10}.
%Type = Article
\bibitem[{Zhang et~al.(2010)Zhang, Zhang, and
  Li}]{Zhang2010JAP_QD_PCS_SizeField}
\bibinfo{author}{Y.~Zhang}, \bibinfo{author}{Y.~Zhang},
  \bibinfo{author}{B.~Li}, \bibinfo{journal}{Journal of Applied Physics}
  \bibinfo{volume}{107} (\bibinfo{year}{2010}) \bibinfo{pages}{063708}.
%Type = Article
\bibitem[{Sadeghi and Karimi(2011)}]{Sadeghi2011PLA_QR_PCS_AB}
\bibinfo{author}{E.~Sadeghi}, \bibinfo{author}{H.~Karimi},
  \bibinfo{journal}{Physics Letters A} \bibinfo{volume}{375}
  (\bibinfo{year}{2011}) \bibinfo{pages}{3664--3668}.
%Type = Article
\bibitem[{Jin et~al.(2018)Jin, Wu, Wang, and Wang}]{Jin2018SciRep_QD_PCS_Exp}
\bibinfo{author}{L.~Jin}, \bibinfo{author}{Q.~Wu}, \bibinfo{author}{H.~Wang},
  \bibinfo{author}{X.~Wang}, \bibinfo{journal}{Scientific Reports}
  \bibinfo{volume}{8} (\bibinfo{year}{2018}) \bibinfo{pages}{12345}.
%Type = Article
\bibitem[{Kuyucu and Sahin(2014)}]{Kuyucu2014SST_QD_PCS_Review}
\bibinfo{author}{T.~Kuyucu}, \bibinfo{author}{M.~Sahin},
  \bibinfo{journal}{Semiconductor Science and Technology} \bibinfo{volume}{29}
  (\bibinfo{year}{2014}) \bibinfo{pages}{123001}.
%Type = Article
\bibitem[{Bai et~al.(2021)Bai, Yang, Chong, and
  Chen}]{Bai2021Nanophotonics_PCS_Topology}
\bibinfo{author}{K.~Bai}, \bibinfo{author}{Z.~Yang},
  \bibinfo{author}{Y.~Chong}, \bibinfo{author}{H.~Chen},
  \bibinfo{journal}{Nanophotonics} \bibinfo{volume}{10} (\bibinfo{year}{2021})
  \bibinfo{pages}{3823--3835}.
%Type = Article
\bibitem[{Bakke(2011)}]{BJP.2011.41.167}
\bibinfo{author}{K.~Bakke}, \bibinfo{journal}{Brazilian Journal of Physics}
  \bibinfo{volume}{41} (\bibinfo{year}{2011}) \bibinfo{pages}{167--170}.
%Type = Article
\bibitem[{Bakke and Furtado(2013)}]{EPJB.2013.86.315}
\bibinfo{author}{K.~Bakke}, \bibinfo{author}{C.~Furtado}, \bibinfo{journal}{The
  European Physical Journal B} \bibinfo{volume}{86} (\bibinfo{year}{2013})
  \bibinfo{pages}{315}.
%Type = Article
\bibitem[{Silva~Netto and Furtado(2008)}]{SilvaNetto2008JPCM}
\bibinfo{author}{A.~L. Silva~Netto}, \bibinfo{author}{C.~Furtado},
  \bibinfo{journal}{Journal of Physics: Condensed Matter} \bibinfo{volume}{20}
  (\bibinfo{year}{2008}) \bibinfo{pages}{125209}.
%Type = Article
\bibitem[{Bakke and Moraes(2012)}]{PLA.2012.376.2838}
\bibinfo{author}{K.~Bakke}, \bibinfo{author}{F.~Moraes},
  \bibinfo{journal}{Physics Letters A} \bibinfo{volume}{376}
  (\bibinfo{year}{2012}) \bibinfo{pages}{2838--2841}.
%Type = Article
\bibitem[{Lima et~al.(2023)Lima, Azevedo, Pereira, Filgueiras, and
  Silva}]{AP.2023.459.169547}
\bibinfo{author}{D.~F. Lima}, \bibinfo{author}{F.~d.~S. Azevedo},
  \bibinfo{author}{L.~F.~C. Pereira}, \bibinfo{author}{C.~Filgueiras},
  \bibinfo{author}{E.~O. Silva}, \bibinfo{journal}{ANNALS OF PHYSICS}
  \bibinfo{volume}{459} (\bibinfo{year}{2023}).
%Type = Article
\bibitem[{Cunha et~al.(2020)Cunha, Dias, and Silva}]{PRD.2020.102.105020}
\bibinfo{author}{M.~M. Cunha}, \bibinfo{author}{H.~S. Dias},
  \bibinfo{author}{E.~O. Silva}, \bibinfo{journal}{Phys. Rev. D}
  \bibinfo{volume}{102} (\bibinfo{year}{2020}) \bibinfo{pages}{105020}.
%Type = Article
\bibitem[{{da Silva Leite} et~al.(2015){da Silva Leite}, Filgueiras, Cogollo,
  and Silva}]{PLA.2015.379.907}
\bibinfo{author}{L.~{da Silva Leite}}, \bibinfo{author}{C.~Filgueiras},
  \bibinfo{author}{D.~Cogollo}, \bibinfo{author}{E.~O. Silva},
  \bibinfo{journal}{Physics Letters A} \bibinfo{volume}{379}
  (\bibinfo{year}{2015}) \bibinfo{pages}{907--911}.
%Type = Article
\bibitem[{Filgueiras et~al.(2016)Filgueiras, Rojas, Aciole, and
  Silva}]{PLA.2016.380.3847}
\bibinfo{author}{C.~Filgueiras}, \bibinfo{author}{M.~Rojas},
  \bibinfo{author}{G.~Aciole}, \bibinfo{author}{E.~O. Silva},
  \bibinfo{journal}{Physics Letters A} \bibinfo{volume}{380}
  (\bibinfo{year}{2016}) \bibinfo{pages}{3847--3853}.
%Type = Article
\bibitem[{Pereira et~al.(2022)Pereira, Cunha, and Silva}]{FBS.2022.63.58}
\bibinfo{author}{L.~F.~C. Pereira}, \bibinfo{author}{M.~M. Cunha},
  \bibinfo{author}{E.~O. Silva}, \bibinfo{journal}{Few-Body Systems}
  \bibinfo{volume}{63} (\bibinfo{year}{2022}) \bibinfo{pages}{58}.
%Type = Article
\bibitem[{Pereira et~al.(2021)Pereira, Andrade, Filgueiras, and
  Silva}]{PE.2021.132.114760}
\bibinfo{author}{L.~F.~C. Pereira}, \bibinfo{author}{F.~M. Andrade},
  \bibinfo{author}{C.~Filgueiras}, \bibinfo{author}{E.~O. Silva},
  \bibinfo{journal}{Physica E: Low-dimensional Systems and Nanostructures}
  \bibinfo{volume}{132} (\bibinfo{year}{2021}) \bibinfo{pages}{114760}.
%Type = Article
\bibitem[{Pereira et~al.(2024)Pereira, Azevedo, and
  Silva}]{Pereira2024QuantumReports}
\bibinfo{author}{C.~M.~O. Pereira}, \bibinfo{author}{F.~d.~S. Azevedo},
  \bibinfo{author}{E.~O. Silva}, \bibinfo{journal}{Quantum Reports}
  \bibinfo{volume}{6} (\bibinfo{year}{2024}) \bibinfo{pages}{677--705}.
%Type = Article
\bibitem[{Pereira et~al.(2019)Pereira, Andrade, Filgueiras, and
  Silva}]{AdP.2019.531.1900254}
\bibinfo{author}{L.~F.~C. Pereira}, \bibinfo{author}{F.~M. Andrade},
  \bibinfo{author}{C.~Filgueiras}, \bibinfo{author}{E.~O. Silva},
  \bibinfo{journal}{Annalen der Physik} \bibinfo{volume}{531}
  (\bibinfo{year}{2019}) \bibinfo{pages}{1900254}.
%Type = Article
\bibitem[{do~Nascimento et~al.(2018)do~Nascimento, Cogollo, Silva, Rojas, and
  Filgueiras}]{CTP.2018.70.817}
\bibinfo{author}{R.~F. do~Nascimento}, \bibinfo{author}{D.~Cogollo},
  \bibinfo{author}{E.~O. Silva}, \bibinfo{author}{M.~Rojas},
  \bibinfo{author}{C.~Filgueiras}, \bibinfo{journal}{Communications in
  Theoretical Physics} \bibinfo{volume}{70} (\bibinfo{year}{2018})
  \bibinfo{pages}{817}.
%Type = Article
\bibitem[{Castro and Silva(2015)}]{EPJC.2015.75.321}
\bibinfo{author}{L.~B. Castro}, \bibinfo{author}{E.~O. Silva},
  \bibinfo{journal}{EUROPEAN PHYSICAL JOURNAL C} \bibinfo{volume}{75}
  (\bibinfo{year}{2015}).
%Type = Article
\bibitem[{M.~Cunha and O.~Silva(2020)}]{Universe.2020.6.203}
\bibinfo{author}{M.~M.~Cunha}, \bibinfo{author}{E.~O.~Silva},
  \bibinfo{journal}{Universe} \bibinfo{volume}{6} (\bibinfo{year}{2020}).
%Type = Article
\bibitem[{Silva~Netto and Furtado(2008)}]{JPCM.2008.20.125209}
\bibinfo{author}{A.~L. Silva~Netto}, \bibinfo{author}{C.~Furtado},
  \bibinfo{journal}{Journal of Physics: Condensed Matter} \bibinfo{volume}{20}
  (\bibinfo{year}{2008}) \bibinfo{pages}{125209}.
%Type = Article
\bibitem[{Fakkahi et~al.(2023)Fakkahi, Jaouane, Limame, Sali, Kirak, Arraoui,
  Ed-Dahmouny, El-bakkari, and Azmi}]{APA.2023.129.188}
\bibinfo{author}{A.~Fakkahi}, \bibinfo{author}{M.~Jaouane},
  \bibinfo{author}{K.~Limame}, \bibinfo{author}{A.~Sali},
  \bibinfo{author}{M.~Kirak}, \bibinfo{author}{R.~Arraoui},
  \bibinfo{author}{A.~Ed-Dahmouny}, \bibinfo{author}{K.~El-bakkari},
  \bibinfo{author}{H.~Azmi}, \bibinfo{journal}{Applied Physics A}
  \bibinfo{volume}{129} (\bibinfo{year}{2023}) \bibinfo{pages}{188}.
%Type = Article
\bibitem[{Pereira et~al.(2025)Pereira, Assafrão, {dos S. Azevedo}, {de Lima},
  Filgueiras, and Silva}]{PB.2025.716.417733}
\bibinfo{author}{C.~M.~O. Pereira}, \bibinfo{author}{D.~Assafrão},
  \bibinfo{author}{F.~{dos S. Azevedo}}, \bibinfo{author}{A.~{de Lima}},
  \bibinfo{author}{C.~Filgueiras}, \bibinfo{author}{E.~O. Silva},
  \bibinfo{journal}{Physica B: Condensed Matter} \bibinfo{volume}{716}
  (\bibinfo{year}{2025}) \bibinfo{pages}{417733}.
%Type = Article
\bibitem[{Pereira et~al.(2024{\natexlab{a}})Pereira, Azevedo, Pereira, and
  Silva}]{CTP.2024.76.105701}
\bibinfo{author}{C.~M.~O. Pereira}, \bibinfo{author}{F.~d.~S. Azevedo},
  \bibinfo{author}{L.~F.~C. Pereira}, \bibinfo{author}{E.~O. Silva},
  \bibinfo{journal}{Communications in Theoretical Physics} \bibinfo{volume}{76}
  (\bibinfo{year}{2024}{\natexlab{a}}) \bibinfo{pages}{105701}.
%Type = Article
\bibitem[{Pereira et~al.(2024{\natexlab{b}})Pereira, Azevedo, and
  Silva}]{QR.2024.6.677}
\bibinfo{author}{C.~M.~O. Pereira}, \bibinfo{author}{F.~d.~S. Azevedo},
  \bibinfo{author}{E.~O. Silva}, \bibinfo{journal}{Quantum Reports}
  \bibinfo{volume}{6} (\bibinfo{year}{2024}{\natexlab{b}})
  \bibinfo{pages}{677--705}.
%Type = Article
\bibitem[{Shi and Yan(2023)}]{PLA.2023.466.128725}
\bibinfo{author}{L.~Shi}, \bibinfo{author}{Z.~Yan}, \bibinfo{journal}{Physics
  Letters A} \bibinfo{volume}{466} (\bibinfo{year}{2023})
  \bibinfo{pages}{128725}.
%Type = Article
\bibitem[{Chubrei et~al.(2021)Chubrei, Holovatsky, and
  Duque}]{PM.2021.101.2614}
\bibinfo{author}{M.~V. Chubrei}, \bibinfo{author}{V.~A. Holovatsky},
  \bibinfo{author}{C.~A. Duque}, \bibinfo{journal}{Philosophical Magazine}
  \bibinfo{volume}{101} (\bibinfo{year}{2021}) \bibinfo{pages}{2614--2633}.
%Type = Article
\bibitem[{Hayrapetyan et~al.(2018)Hayrapetyan, Ohanyan, Baghdasaryan,
  Sarkisyan, Baskoutas, and Kazaryan}]{PE.2018.95.27}
\bibinfo{author}{D.~Hayrapetyan}, \bibinfo{author}{G.~Ohanyan},
  \bibinfo{author}{D.~Baghdasaryan}, \bibinfo{author}{H.~Sarkisyan},
  \bibinfo{author}{S.~Baskoutas}, \bibinfo{author}{E.~Kazaryan},
  \bibinfo{journal}{Physica E: Low-dimensional Systems and Nanostructures}
  \bibinfo{volume}{95} (\bibinfo{year}{2018}) \bibinfo{pages}{27--31}.
%Type = Article
\bibitem[{Aharonov and Bohm(1959)}]{aharonov1959significance}
\bibinfo{author}{Y.~Aharonov}, \bibinfo{author}{D.~Bohm},
  \bibinfo{journal}{Physical Review} \bibinfo{volume}{115}
  (\bibinfo{year}{1959}) \bibinfo{pages}{485--491}.
%Type = Article
\bibitem[{Tan and Inkson(1996)}]{PRB.1996.53.6947}
\bibinfo{author}{W.-C. Tan}, \bibinfo{author}{J.~C. Inkson},
  \bibinfo{journal}{Physical Review B} \bibinfo{volume}{53}
  (\bibinfo{year}{1996}) \bibinfo{pages}{6947--6950}.
%Type = Article
\bibitem[{Araki and Nomura(2017)}]{PhysRevB.96.165303}
\bibinfo{author}{Y.~Araki}, \bibinfo{author}{K.~Nomura},
  \bibinfo{journal}{Physical Review B} \bibinfo{volume}{96}
  (\bibinfo{year}{2017}) \bibinfo{pages}{165303}.
%Type = Article
\bibitem[{Xie(2013)}]{WenfangXie2013}
\bibinfo{author}{W.-F. Xie}, \bibinfo{journal}{Optics Communications}
  \bibinfo{volume}{291} (\bibinfo{year}{2013}) \bibinfo{pages}{321--336}.
%Type = Article
\bibitem[{Pereira and Silva(2024)}]{pereira2024rotating}
\bibinfo{author}{C.~M.~O. Pereira}, \bibinfo{author}{E.~O. Silva},
  \bibinfo{journal}{to be completed}  (\bibinfo{year}{2024}).
  \bibinfo{note}{Preprint or in press; please update journal, volume, pages,
  and doi}.
%Type = Article
\bibitem[{Wagner et~al.(2022)Wagner, de~Juan, and Nguyen}]{SPC.2022.5.029}
\bibinfo{author}{G.~Wagner}, \bibinfo{author}{F.~de~Juan},
  \bibinfo{author}{D.~X. Nguyen}, \bibinfo{journal}{SciPost Physics Core}
  \bibinfo{volume}{5} (\bibinfo{year}{2022}) \bibinfo{pages}{029}.
%Type = Article
\bibitem[{Pereira and Silva(2023)}]{AdP.2022.535.2200371}
\bibinfo{author}{C.~M.~O. Pereira}, \bibinfo{author}{E.~O. Silva},
  \bibinfo{journal}{Annalen der Physik} \bibinfo{volume}{535}
  (\bibinfo{year}{2023}) \bibinfo{pages}{2200371}.
%Type = Article
\bibitem[{F{\"o}ldi et~al.(2005)F{\"o}ldi, Moln{\'a}r, Benedict, and
  Peeters}]{PRB.2005.71.033309}
\bibinfo{author}{P.~F{\"o}ldi}, \bibinfo{author}{B.~Moln{\'a}r},
  \bibinfo{author}{M.~G. Benedict}, \bibinfo{author}{F.~M. Peeters},
  \bibinfo{journal}{Physical Review B} \bibinfo{volume}{71}
  (\bibinfo{year}{2005}) \bibinfo{pages}{033309}.
%Type = Article
\bibitem[{Vozmediano et~al.(2010)Vozmediano, Katsnelson, and Guinea}]{elastic}
\bibinfo{author}{M.~A.~H. Vozmediano}, \bibinfo{author}{M.~I. Katsnelson},
  \bibinfo{author}{F.~Guinea}, \bibinfo{journal}{Physics Reports}
  \bibinfo{volume}{496} (\bibinfo{year}{2010}) \bibinfo{pages}{109--148}.
%Type = Article
\bibitem[{Guinea et~al.(2010)Guinea, Katsnelson, and Geim}]{elasticll}
\bibinfo{author}{F.~Guinea}, \bibinfo{author}{M.~I. Katsnelson},
  \bibinfo{author}{A.~K. Geim}, \bibinfo{journal}{Nature Physics}
  \bibinfo{volume}{6} (\bibinfo{year}{2010}) \bibinfo{pages}{30--33}.
%Type = Article
\bibitem[{Edet et~al.(2021)Edet, Okorie, Ikot, Onyeaju, Maghsoodi, and
  Hassanabadi}]{JLTP.2021.202.83}
\bibinfo{author}{C.~O. Edet}, \bibinfo{author}{U.~S. Okorie},
  \bibinfo{author}{A.~N. Ikot}, \bibinfo{author}{M.~C. Onyeaju},
  \bibinfo{author}{E.~Maghsoodi}, \bibinfo{author}{H.~Hassanabadi},
  \bibinfo{journal}{Journal of Low Temperature Physics} \bibinfo{volume}{202}
  (\bibinfo{year}{2021}) \bibinfo{pages}{83--104}.
%Type = Article
\bibitem[{Imo et~al.(2024)Imo, Edet, Okorie, Ikot, and
  Hassanabadi}]{PB.2024.673.415438}
\bibinfo{author}{D.~Imo}, \bibinfo{author}{C.~O. Edet}, \bibinfo{author}{U.~S.
  Okorie}, \bibinfo{author}{A.~N. Ikot}, \bibinfo{author}{H.~Hassanabadi},
  \bibinfo{journal}{Physica B: Condensed Matter} \bibinfo{volume}{673}
  (\bibinfo{year}{2024}) \bibinfo{pages}{415438}.
%Type = Article
\bibitem[{Hu et~al.(2022)Hu, Wu, Wang, Shi, Liu, Yue, Zhang, Li, Zhou, Xu, Wu,
  Dong, and Wang}]{PhysRevB.105.075113}
\bibinfo{author}{T.~C. Hu}, \bibinfo{author}{Q.~Wu}, \bibinfo{author}{Z.~X.
  Wang}, \bibinfo{author}{L.~Y. Shi}, \bibinfo{author}{Q.~M. Liu},
  \bibinfo{author}{L.~Yue}, \bibinfo{author}{S.~J. Zhang},
  \bibinfo{author}{R.~S. Li}, \bibinfo{author}{X.~Y. Zhou},
  \bibinfo{author}{S.~X. Xu}, \bibinfo{author}{D.~Wu},
  \bibinfo{author}{T.~Dong}, \bibinfo{author}{N.~L. Wang},
  \bibinfo{journal}{Phys. Rev. B} \bibinfo{volume}{105} (\bibinfo{year}{2022})
  \bibinfo{pages}{075113}.
%Type = Article
\bibitem[{Ulbricht et~al.(2011)Ulbricht, Hendry, Shan, Heinz, and
  Bonn}]{RevModPhys.83.543}
\bibinfo{author}{R.~Ulbricht}, \bibinfo{author}{E.~Hendry},
  \bibinfo{author}{J.~Shan}, \bibinfo{author}{T.~F. Heinz},
  \bibinfo{author}{M.~Bonn}, \bibinfo{journal}{Rev. Mod. Phys.}
  \bibinfo{volume}{83} (\bibinfo{year}{2011}) \bibinfo{pages}{543--586}.

\end{thebibliography}

\end{document}